\documentclass[12pt]{article}
\usepackage{graphicx,amsmath,amssymb}
\usepackage{color}

\usepackage{multirow}
\usepackage{subfigure}
\usepackage{setspace}

\newlength{\dinwidth}
\newlength{\dinmargin}
\def\be{\begin{equation}}   
\def\ee{\end{equation}}  
\def\bea{\begin{eqnarray}}                      
\def\eea{\end{eqnarray}}
\def\ch1{$\chi(1^+)$}
\def\lapproxeq{\lower .7ex\hbox{$\;\stackrel{\textstyle                                                    
<}{\sim}\;$}}                                                    
\def\gapproxeq{\lower .7ex\hbox{$\;\stackrel{\textstyle                                                    
>}{\sim}\;$}}

\begin{document}

\begin{flushright}                                                    
IPPP/17/78 \\
\today \\                                                    
\end{flushright} 

\vspace*{0.5cm}

\vspace*{0.5cm}
\begin{center}
{\Large \bf Multiple interactions and rapidity gap survival} \\ 
\vspace*{1cm}

V.A. Khoze$^{a,b}$, A.D. Martin$^{a}$ and M.G. Ryskin$^{a,b}$\\

\vspace*{0.5cm}
$^a$ Institute for Particle Physics Phenomenology, Durham University, 
Durham, DH1 3LE, UK\\ 

$^b$ Petersburg Nuclear Physics Institute, NRC `Kurchatov Institute', 
Gatchina, St.~Petersburg, 188300, Russia \\
\end{center}

\begin{abstract}
Observations of rare processes containing large rapidity gaps at high energy colliders may be exceptionally informative. However the cross sections of these events are small in comparison with that for the inclusive  processes since there is a large probability that the gaps may be filled by secondary particles arising from additional soft interactions or from gluon radiation. Here we review the calculations of the probability that the gaps survive population by particles from these effects for a wide range of different processes.

\end{abstract}
\vspace*{0.5cm}
\singlespacing
{\small
\tableofcontents
  }
\section{Introduction}
Processes containing rapidity gaps\footnote{A rapidity gap is an interval in the rapidity variable, $y$, which contains no particles. Rapidity, $y\equiv \frac{1}{2} {\rm log}[(E+p_L)/(E-p_L)]$, is an alternative way to express the longitudinal momentum $p_L$ of a particle of energy $E$. Unlike $p_L$, it has the advantage of being {\it additive} under Lorentz boosts along the axis of motion. In the c.m. frame of a $pp$ collision of high energy $\sqrt{s}$, a particle of mass $M$ lies in the rapidity interval $\pm{\rm log}(\sqrt{s} /M)$. Actually the experiments usually use pseudorapidity $\eta=-{\rm log(tan}(\theta/2))$ where $\theta$ is the polar angle of the outgoing particle. For a massless particle, or for $E\gg M$, we have $\eta=y$.}, which may be observed at the LHC,  are potentially very informative.
 An example of one class of such reactions is the central exclusive process $pp\to p+X+p$ where a massive system $X$ is produced by either gluon-gluon, or $\gamma\gamma$ or $W^+W^-$ fusion with a large rapidity gap on either side.  With the advent of very forward proton detectors such exclusive processes have the potential of discovering new Beyond the Standard Model (BSM) Physics.  However, the survival of the rapidity gap to population by other particles in the experiments at the LHC and at other particle colliders needs to be taken carefully into account.  It is a challenging problem both experimentally and theoretically.  In this review we focus on the calculation of the probability of the survival of the rapidity gap for a range of different processes.

As a rule, theoretically we calculate  {\em inclusive} cross sections
which include all possible configurations of the final particles without any
additional restrictions. On the other hand, in a real experiment usually a  specific class of events is selected. To suppress the background we have to introduce some {\em trigger} conditions and experimental {\em cuts}.

Actually we never deal  with a {\em completely inclusive} process, so the original theoretical result must be corrected to account for this fact.
In some cases, in particular in high-energy, low-multiplicity exclusive reactions, the corresponding corrections are quite large.

Indeed, in proton-proton collisions there is a large probability for additional soft interactions of the spectator partons that accompany the partons of the main subprocess of interest. Let us say that the mean number of these Multiple Interactions (MI) is $\langle n \rangle$. These MI produce a large number of new secondaries which may violate the 
original conditions (cuts) that were imposed to select the events of interest. Thus we have to consider only the rare events where there are no such  additional interactions. For example, if for simplicity, we assume a Poisson distribution then the correction factor, which is equal to the probability to have no inelastic interaction, is $P_0=\exp(-\langle n \rangle)$, -- that is the probability is exponentially suppressed.  The exact form of the suppression will be derived later.

The most popular condition used to select an exclusive events is a pseudorapidity cut. That is only events without any hadronic activity (or the events without any jets with a relatively large transverse energy $E_T>E_0$) in some large rapidity interval are selected.
In this case we have to calculate the probability that an additional interaction will not produce secondaries which populate the rapidity gap. This probability is called the gap survival factor, $S^2$ \cite{Bjorken:1991xr}.

\subsection{Gap survival factor}
The problem of {\it gap survival} was raised for the first time in~\cite{Dokshitzer:1987nc,Dokshitzer:1991he} where the process of Higgs boson production via $WW$-fusion was considered. To study this process it was proposed to select events with large gaps in the rapidity intervals corresponding to $W$-boson exchange, since unlike gluon exchange, a colourless $W$-boson does not produce any accompanying secondary particles.
However the cross section of such  events is considerably suppressed by an $S^2$ factor arising from the MI interactions of the spectators of the colliding protons.  The corresponding $S^2$ factor was calculated using the PYTHIA Monte Carlo \cite{Bengtsson:1987kr} in~\cite{Dokshitzer:1991he}.  

The first analytical evaluation of a survival factor was performed by Bjorken~\cite{Bjorken:1991xr}. Subsequently, 
the values of $S^2$ expected at LHC energies have been calculated by many groups
using various models for the high-energy proton-proton interaction, see, for example, ~\cite{KMR}~-~\cite{Fagundes:2017xli}. 
Strictly speaking the value of the $S^2$ factor {\em depends} on the particular process and the cuts imposed in the experiment. Originally the bulk of analytic calculations of the $S^2$ were performed in the context
of pure exclusive Higgs boson production 
\begin{equation}
pp\to p~+~H~+~p\ ,
\end{equation}
where the plus signs
 indicate the presence of Large Rapidity Gaps (LRG) either side of the produced Higgs boson.
 In spite of the small cross section, such a process with a LRG has an important advantage of allowing the study of the Higgs boson, or any new BSM particle, in a very clean experimental environment,
see e.g. \cite{Khoze:2001xm}~-~\cite{N.Cartiglia:2015gve}.
 Here any additional particle violates the exclusivity, so the theoretical calculation of $S^2$ is well-defined and straightforward.
 
 Another possibility is to use a general purpose Monte Carlo (MC) generator
which includes a MI option. This may allow a better account of the real experimental selection. However the approach used to implement the {\em soft} MI option in a MC is usually simplified and less precise than that in the analytic calculations. The problem is that we do not have a good theory for high-energy soft interactions written in terms of quarks and gluons -- the QCD degrees of freedom used in the majority of
the MC generators. The only possibility is to use some phenomenological model with parameters tuned to reproduce the data on the elastic $pp$-cross section ($d\sigma/dt$) and the cross sections of diffractive proton dissociation.
Such a phenomenological model is easier to implement analytically.
 An exception is the QGSJET MC \cite{Ostapchenko:2010gt,Ostapchenko:2013pia},
 where the most complete soft Reggeon theory is implemented.  The various Monte Carlos which account for rapidity gap events are briefly discussed in Section \ref{sec:9}.
 
 Before considering gap survival phenomena in more detail, we emphasize that the value of $S^2$ is not universal. First of all
 $S^2(b)$ depends on the impact parameter $b$ -- the separation of the two incoming protons in transverse plane. At larger $b$ the probability of an additional interaction is smaller.
 Of course experimentally we cannot fix the value of $b$. In each
particular process the amplitude is given by an integral over $b$. Thus, 
depending on the transverse momenta, and the detailed structure of the
matrix element, the final contribution comes from  smaller or larger
impact parameters.

For example, if instead of a Higgs boson of spin, parity $J^P=0^+$, we were to consider the Central Exclusive Production (CEP) of a pseudo-scalar ($J^P=0^-$) boson of the Minimal Supersymmetric Higgs sector (see e.g.\cite {Heinemeyer:2007tu}),
then the amplitude will contain an antisymmetric tensor $\epsilon^{\mu\nu\alpha\beta}$ which cannot be saturated at $b=0$; there are not enough vectors in transverse plane. The vanishing of the amplitude at $b=0$ leads to a larger mean impact parameter, $b$, and correspondingly to a larger gap survival 
probability $\langle S^2\rangle $ (see e.g.~\cite{Kaidalov:2003fw}).
Thus one cannot use the same value of $S^2$ for different CEP processes. $S^2$ must be calculated in each case individually.

\subsection{Factorization breaking  \label{sec:1.2}}
Recall that in general an additional MI breaks both collinear and high-energy (or $k_T$) factorization.  The well known example is
  high $E_T$ diffractive jet 
production at the Tevatron~\cite{Affolder:2000vb}.
 Based on factorization, the corresponding cross section should be calculated as the convolution of the parton distributions in the proton and in the Pomeron (which transfers the momentum across the gap without producing any new secondaries) with the `hard' matrix element, $M$, which describes the high $E_T$ jet formation
\be
\label{f-br}
\sigma={\rm PDF}_{\rm proton}(x_1)\otimes |M|^2\otimes {\rm PDF}_{\rm pomeron}(x_2)\ .
\ee   
Both parton distributions are known from HERA electron-proton collider data on Deep Inelastic Scattering (DIS) and on
diffractive DIS.

\begin{figure}
\begin{center}
\vspace{-1cm}
\includegraphics[height=8cm]{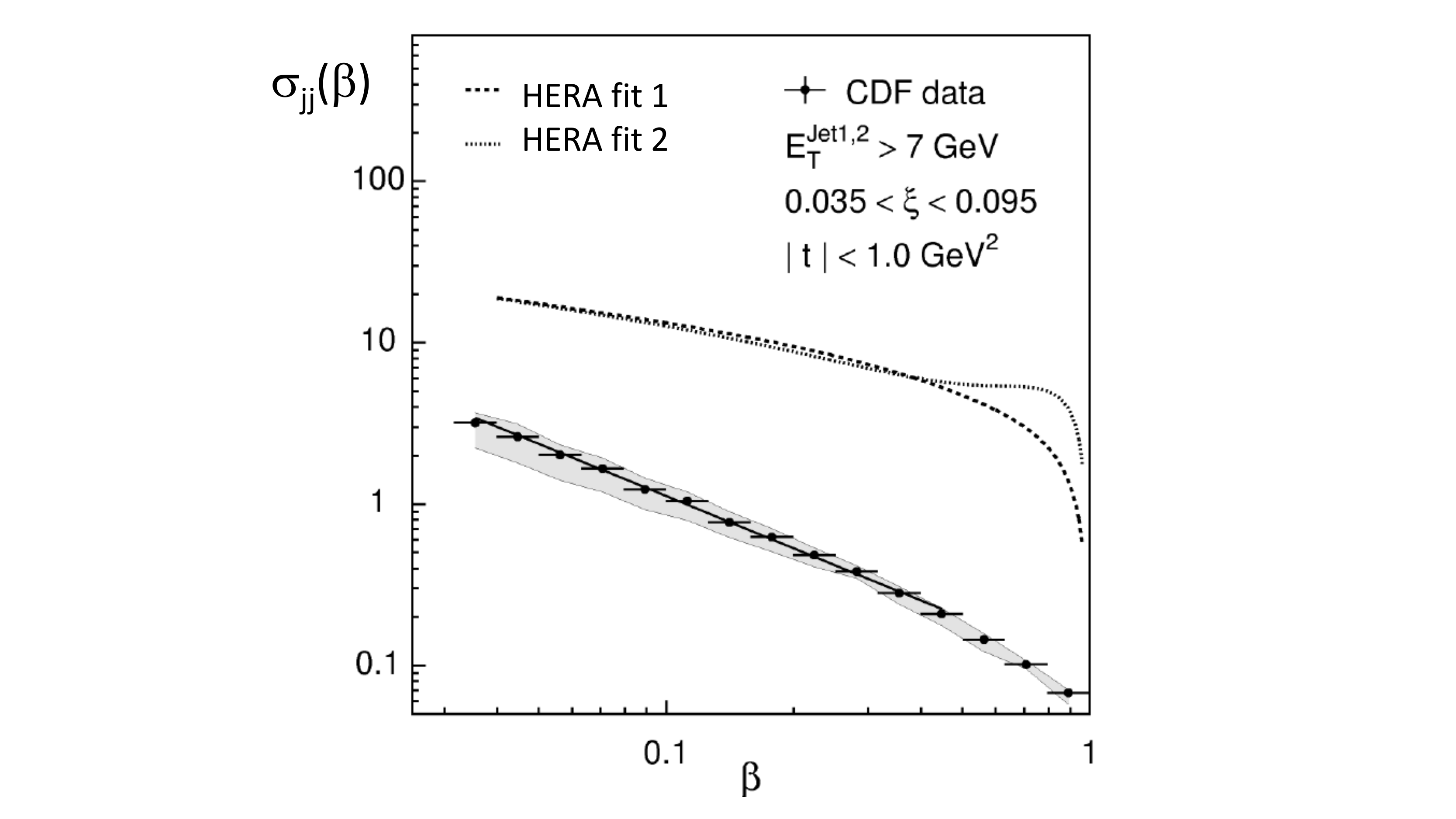}
\vspace{0cm}
\caption{The $\beta$ dependence of the diffractive dijet cross section, $\sigma_{jj}$, compared with predictions based on factorization and the diffractive PDFs measured by the H1 collaboration at HERA.  The figure is adapted from \cite{Affolder:2000vb}.}
\label{fig:S2a}
\end{center}
\end{figure}

\begin{figure}
\begin{center}
\vspace{-1cm}
\includegraphics[height=8cm]{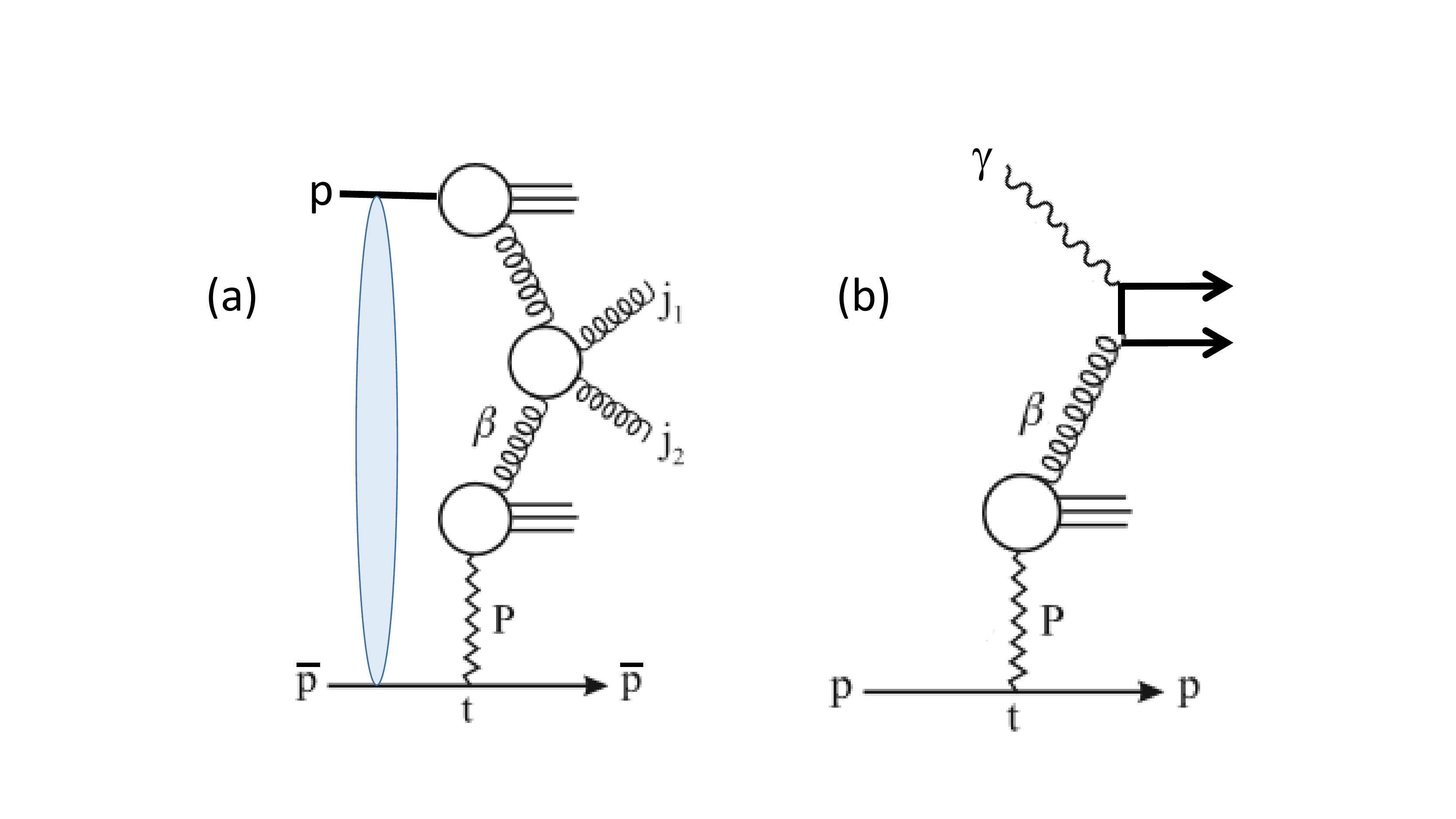}
\vspace{0cm}
\caption{Schematic diagrams for (a) Diffractive dijet production in $p\bar{p}$ collisions; (b) diffractive deep inelastic scattering at HERA. The rapidity gap indicated by the pomeron, P, is common to both diagrams.}
\label{fig:1ab}
\end{center}
\end{figure}

However we see from Fig.~\ref{fig:S2a} that the  cross section measured by the CDF collaboration
 turns out to be almost an order of magnitude
 smaller than that given by (\ref{f-br}). As explained in \cite{Kaidalov:2001iz}, in this proton-antiproton collision the cross section was suppressed by an $S^2$ factor whose origin is indicated by the ellipse in Fig.~\ref{fig:1ab}(a). Recall that for diffractive DIS of Fig.~\ref{fig:1ab}(b) the probability of an additional interaction of the heavy incident photon is small. 
  
 Moreover, already in this process, it was observed that the $S^2$  suppression is not a constant but depends on the pomeron momentum fraction, $\beta$, carried by the  dijet system. 
That is in the different kinematic regions the probability of the survival of the LRG is different; see also the end of subsection~\ref{sec:8.1}.

More recently a similar  $S^2$  suppression has been observed in the high $E_T$ diffractive dijet production 
in the  $\sqrt{s}=$ 7 TeV $pp$ collisions at the LHC by the ATLAS~\cite{Aad:2015xis} and CMS~\cite{Chatrchyan:2012vc,Sirunyan:2017rdp} collaborations.
Factorization breaking effects were also intensively studied in  diffractive dijet photoproduction
at HERA \cite{ValkaRova:2015vpa}~-~\cite{Zlebcik:2011kq}.

\section{Physical origins of the survival factor, $S^2$   \label{sec:2}}
The processes which violate exclusive or LRG kinematics may have different origins. There may be a soft inelastic interaction which produces the new secondary particles which populate the gap (eikonal suppression) or hard QCD bremsstrahlung (Sudakov suppression) or rescattering of the leading hadron (migration). In this section we  describe these three main effects in turn.

\subsection{Soft inelastic interactions and $S^2_{\rm eikonal}$   \label{sec:2.1}}
In the simplest approximation the probability to have {\em no} 
 additional soft interaction in the proton-proton collision at fixed impact parameter, $b$, is given by the factor 
\be
 \label{S2b}
 S^2(b)=\exp(-\Omega(b))
\ee
  where
the proton-proton opacity (optical density) $\Omega(b)$ can be `measured'
via elastic scattering. Indeed, within the one-channel eikonal model
the amplitude in $b$ space reads
\begin{equation}
\label{elamp}
T_{\rm el}(b)~=~1-e^{-\Omega(b)/2},
\end{equation}
while the probability of an inelastic interaction
\begin{equation}
G_{\rm inel}(b)~=~1-e^{-\Omega(b)}\ .
\end{equation}
That is $\exp(-\Omega(b)$ is the probability to have no inelastic interaction, and hence (\ref{S2b}).

 Recall that (\ref{elamp}) is a very general expression. It is the solution of
 the $s$-channel elastic unitarity equation\footnote{The unitarity relation is ${\cal SS}^\dagger =I$ where ${\cal S}$ is the scattering matrix. Eq. (\ref{un}) follows from ${\cal S}=I+iT$. }
 \be\label{un}
 2 \mbox{Im} T_{\rm el}(s,b)~=~|T_{\rm el}(s,b)|^2+G_{\rm inel}(s,b)
 \ee
where $G_{\rm inel}$ is the sum over all inelastic intermediate states. In the momentum space the amplitude 
\begin{equation}
T_{\rm el}(q_t)=2i\int d^2b ~T_{\rm el}(b)~e^{-iq_t\cdot  b}
\end{equation}
and the elastic cross section is equal to
\begin{equation}
\label{eik1}
\frac{d\sigma_{\rm el}}{dt}~=~\frac{|T_{\rm el}(t)|^2}{16\pi}
\end{equation}
where $t=-q_t^2$.
We see from (\ref{eik1}) that the elastic scattering data measure the amplitude $T(b)$ and the opacity $\Omega(b)$, if we neglect the small real part of $T(t)$.

However, in this simplified approach we have neglected the possibility of diffractive excitation in 
the intermediate state. It is possible that, after the first interaction, the leading proton will form a heavier nucleon state, like $N^*$, which after the next interaction (or central exclusive process (CEP)) will produce the leading proton back, $p\to N^*\to p$ (or a longer more complicated chain).
To account for these possibilities we have to consider a multi-channel eikonal model. This can be done by implementing the Good-Walker formalism~\cite{GW}. Quantitatively, the contribution caused by proton excitation can be evaluated from data on low-mass single proton dissociation,
$pp\to p+X$. As a rule the models with just two Good-Walker eigenstates are used. That is we consider the proton and one 
nucleon resonance $N^*$ which effectively reproduces the contribution of
all possible excitations ~\cite{Khoze:2013dha,GLM,Ostapchenko:2010gt}.  

A brief introduction to the Good-Walker formalism is given in subsection~\ref{sec:4.4}.  In fact the subject of the lengthy subsection~\ref{sec:4.4} is the calculation of the $S^2_{\rm eik}$ suppression of the exclusive process $pp \to p+X+p$ where here $X$ is a massive object.  A detailed description of the `eikonal' gap survival suppression factor $S^2_{\rm eik}$ can also be found, for example, in \cite{Khoze:2013dha,KMRsoft,Ryskin:2009tk}.

\subsubsection{Enhanced eikonal \label{sec:2.1.1}}
Note that besides an additional interaction between the two incoming protons there may be an interaction of one proton with some intermediate partons
in the hemisphere of the other proton, hidden in the diagram which describes the main CEP process, see the $S^2_{\rm enh}$ ellipse in Fig.~\ref{fig:pCp}. As far as we consider a high scale CEP reaction, such as
 production of a heavy Higgs (or a new BSM) boson or a high $E_T$ dijet, the corresponding effect is rather weak, see subsection~\ref{sec:4.5}.

A weak $S_{\rm enh}$ suppression for large $M_X$ arises because the intermediate partons have a relatively large transverse momenta, $k_t$, so that the cross section,
$\sigma\propto 1/k^2_t$ is small. However, for central production of moderate mass (say,
 $\chi_c$-charmonium) or for soft processes with a LRG, there may be a noticeable additional suppression. In terms of Regge theory
 these effects are described by so-called enhanced (semi-enhanced)\footnote{The effect is {\it enhanced} by a large number of intermediate partons.} Pomeron diagrams. It was most consistently calculated by Ostapchenko in~\cite{Ostapchenko:2010gt,Ostapchenko:2013pia}.

\subsubsection{Observation of central exclusive production at the LHC   \label{sec:2.1.2}} 
Up to now we have considered only purely exclusive reactions, such as (1).
These processes can be selected using forward proton spectrometers.
Dedicated new forward detectors have been installed at the LHC
by both the ATLAS (ATLAS Forward Proton detector \cite{AFP})
and the CMS (CT-PPS detector \cite{CT-PPS}) collaborations; data taking started in June 2016.

A wealth of physics is expected to come from the 2017 high luminosity run.
The purpose of these near-beam
detectors is to measure intact protons arising 
at small angles, giving access to
a wide range of CEP processes. This, in particular, 
can 
 provide a new window
on physics beyond the Standard Model (SM) at the LHC, see, for example, \cite{N.Cartiglia:2015gve,Royon1}.
The forward detectors are installed at about 200 metres on both sides of  the CMS and ATLAS central detectors.  Their coverage in the fractional momentum loss, $\xi$, of the intact protons is optimistically about
$0.015 < \xi < 0.15$ at nominal accelerator and beam conditions, which corresponds
to an acceptance of 200 to 2000 GeV in the invariant mass of the central system 
 when both protons stay intact. Detection of the outgoing
protons allows strong background suppression and kinematic constraints, opening
a promising way to search for new particles and for deviations from the SM.

  By studying a process with a LRG with the help of the {\em veto} trigger in a fixed rapidity interval we actually allow for the contribution  of events with some additional hadron activity in regions not covered by the `veto'. This means that when calculating the suppression factor $S^2(b)=\exp(-\Omega(b))$ we do not have to use the full
 opacity $\Omega(b)$ but only the part which, in terms of 
 $G_{\rm inel}(b)=1-\exp(-\Omega(b))$, corresponds to the processes where the new secondaries are produced within the vetoed interval.

 \subsection{Sudakov suppression   \label{sec:2.2}}
 In the case of QCD-induced CEP the central object is actually formed by the fusion of two coloured particles. For example the Higgs boson is formed by gluon-gluon fusion ($gg\to H$); while the colour flow across the LRG is {\it screened} at larger distances by the exchange of another, relatively soft, $t$-channel gluon~\cite{BL,KMRh}. 
 In the normal situation the `annihilation' of two energetic coloured particles into a colourless object must be accompanied by strong bremsstrahlung. That is besides the Higgs boson (or any other heavy object) we normally  observe a number of jets (mini-jets) in the central rapidity interval. The probability not to have such  jets, that is to  deal with an exclusive process, is given by a Sudakov form factor $T^2$. We can calculate $T^2$ within the leading logarithmic approximation~\cite{KMRh,KMRH}. To leading log accuracy the number of jets obeys a Poisson distribution and the value of $T^2$ can be written as
 \begin{equation}
 T^2=\exp(-\langle n_{\rm jet}\rangle)\ ,
\end{equation}   
 where $\langle n_{\rm jet}\rangle$ is the mean number of jets, which is given by the argument of the exponential in (\ref{tsp}) below.
 
  The factor $T^2$ depends on the mass of the central object and on the transverse momentum of the screening gluon (i.e. on the typical distance at which the colour flow is screened). Besides this there may be some additional experimental conditions, say,  not all the hadron activity is vetoed but some mini-jets with relatively low $E_T<E_0$ are allowed. This, of course, will enlarge the corresponding value of $T^2$. Therefore usually the $T^2$-factor is included in the main CEP amplitude and not in $S^2$; the gap survival probability in this case accounts for the additional `soft' effects only.
  
  An important role of the $T^2$-factor was demonstrated in \cite{Aaltonen:2007hs},
  where the CEP dijet cross section was measured. If account is taken of the $T^2$ suppression then we obtain a much steeper $E_T$ dependence of the exclusive dijet cross section. Without the Sudakov factor the prediction would
disagree with the data.
The most precise calculation of the
$T^2$ factor was performed in~\cite{TF}, where not only the leading double logarithms but also the single log corrections were included.

The Sudakov effect will be discussed in much more detail in subsection~\ref{sec:4.2}.

  \subsection{Migration  \label{sec:2.3}}     
Using a forward proton detector the existence of a large rapidity  gap (LRG) is provided when the outgoing proton carries away practically all of the initial energy. The proton momentum fraction $x_L=1-\xi$ is then very close to 1 and the remaining part, $\xi$, is too small to produce new particles which have sufficiently high energy to populate the LRG region. In such a case one can work without  `veto' detectors. What is the sense of the gap survival factor
in this situation?

An additional interaction not only produces new secondaries but also changes  the momentum of the final nucleon leading to the `migration'
of the measured (final) proton momentum in the  $x_L$ and $p_t$ space.
The  meaning of $S^2$ in this case is the probability not to change the nucleon momentum via an additional interaction. Note that we can only decrease the value of $x_L$. On the other hand the value of the transverse momentum $p_t$ may be either decreased or increased.
In the region of rather large $p_t$ 
the possibility of rescattering may even {\em enlarge} the observed cross section. Here we have to account not only for the inelastic interaction but also for the elastic rescattering which practically does not change $x_L$ but provides  migration in $p_t$ space.
 
 As discussed in~\cite{KKMR-m,KMR-ln} migration may be important when
  describing the spectra of leading neutrons in the region of not too large $x_L<0.8$.

  \section{QCD-induced processes with a hard scale driven by only one PDF   \label{sec:3}}
There is a class of processes with a LRG where the cross section is given by the convolution of the square of the hard matrix element with {\em one} parton distribution. This may be high $E_T$ jet production in diffractive deep inelastic scattering (DDIS), see Fig.~\ref{fig:1ab}(b), or prompt heavy quark-pair production,
and other similar processes where the target proton is separated by a LRG from the
hadronic system produced by the heavy (or prompt) photon.  That is, here we consider the case where the hadronic system is produced by an object which does not participate in the strong 
interaction. Strictly speaking these are not  exclusive reactions, but processes with a LRG. The corresponding cross sections can be written in factorized form\footnote{Neglecting any other possible experimental cuts.} as the convolution  
\be
|M|^2\otimes {\rm PDF}_{\rm pomeron}\otimes {\rm flux}_{\rm pomeron}
\ee 
 where the product of last two terms can be found from the data on diffractive DIS (i.e. by fitting the value of $F_2$ measured in events with LRG).
 
 In this situation the probability of gap survival is very close to 1, $S^2\simeq 1$. Indeed, the cross section is negligibly small for the heavy photon 
 (or other object which does not participate in the strong 
interaction) with the target proton inelastic interaction; that is  the corresponding opacity $\Omega\to 0$ and 
$S^2=\exp(-\Omega)\to 1$. In terms of Feynman diagrams this does not mean that we have no absorptive corrections at all. There exist diagrams with an additional interaction between the proton and some intermediate partons inside the Pomeron PDF, which, for example, we will see later in Fig.~\ref{fig:SS}(b). However the suppression caused by these diagrams is {\em already included} in the 
Pomeron PDF fitted to the diffractive DIS data. 

The same happens in the pure exclusive case when we calculate, say, diffractive $J/\psi$ production amplitude as the convolution of the proton PDF with the `hard' $\gamma+gg\to J/\psi$ amplitude~\cite{J-psi,jones}. The cross section of the
$J/\psi$-proton interaction is  small while the possible interactions between the target proton and some intermediate partons hidden in the proton PDF (and/or between the intermediate partons themselves) is already included in the phenomenological proton PDF taken from the global parton analyses. 
 Thus this class of exclusive processes with a LRG can be calculated by the factorization approach keeping $S^2=1$.

  \section{QCD CEP driven by two PDFs: $pp \to p+X+p$   \label{sec:4}}
 Here we discuss in detail the prediction for a central exclusive process of the type
 \be
 pp \to p+X+p
 \ee
 where a massive system $X$ is produced with LRGs {\it either} side.  This is an important process since the LHC experiments with forward proton detectors may obtain information on `New Physics' in a clean environment.  Moreover, these processes are amenable to reliable perturbative QCD predictions.  The ingredients of the calculation are shown in Fig.~\ref{fig:pCp}.   We will proceed in stages. We start with the formula for the `bare' CEP amplitude. We then show the importance of embedding the Sudakov suppression in the formalism to ensure infrared convergence.  Finally with consider the eikonal corrections.

 \subsection{The QCD-induced CEP amplitude   \label{sec:4.1}}
 
 \begin{figure}
\begin{center}
\includegraphics[height=7cm]{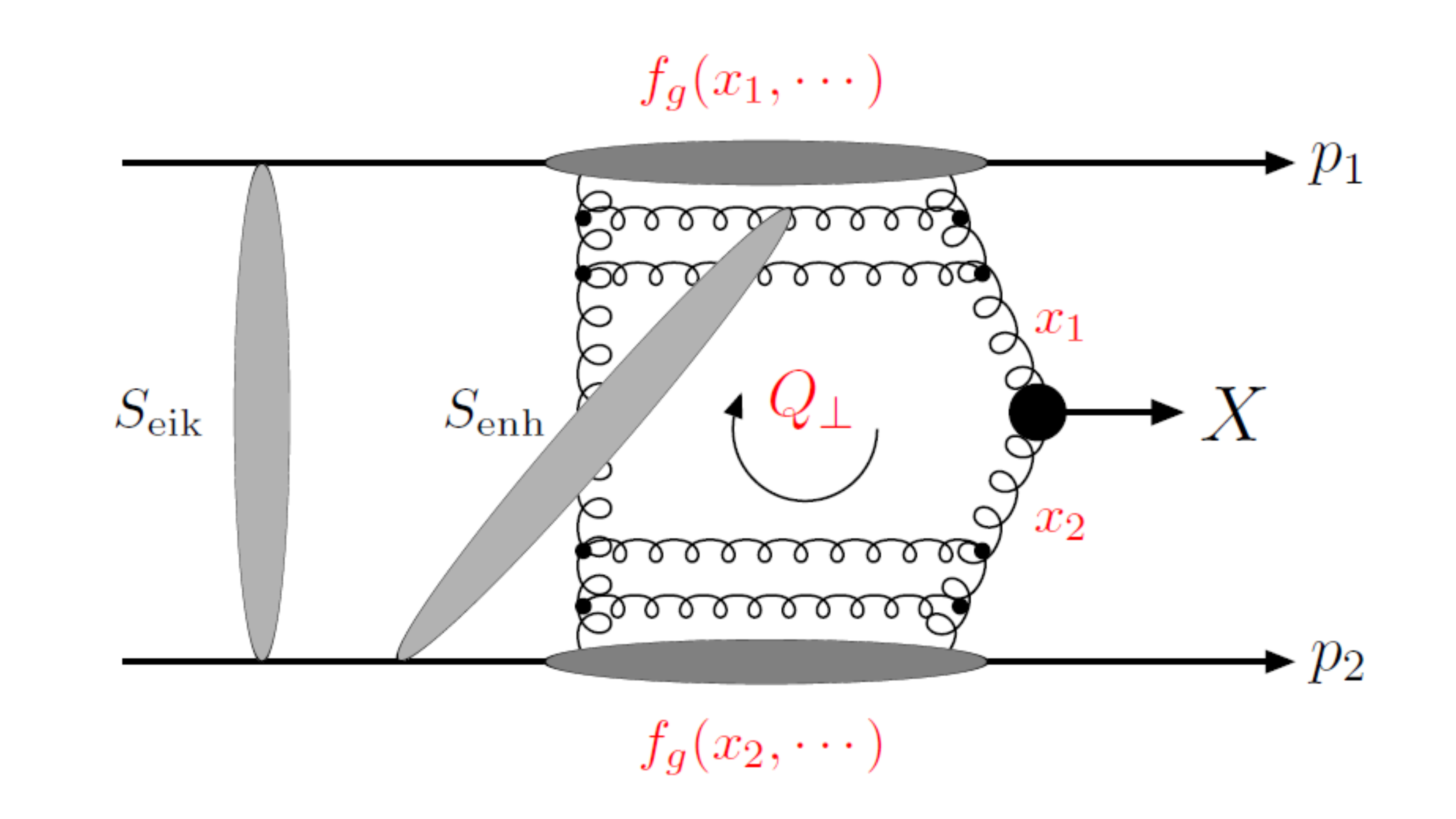}
\vspace{0cm}
\caption{The perturbative mechanism for the exclusive process $pp \to p\,+\, X \, +\, p$, with the eikonal and enhanced survival factors 
shown symbolically.}
\label{fig:pCp}
\end{center}
\end{figure}

CEP processes that proceed purely by the strong interaction can be described by the `Durham' model, a pQCD--based approach that may be applied when the object mass $M_X$ is sufficiently high, see~\cite{Albrow:2010yb}~-~\cite{Harland-Lang:2015eqa} for reviews. The formalism used to calculate the perturbative CEP cross section is explained in detail
 elsewhere~\cite{KMRH,Harland-Lang:2014lxa} and we  only present a very brief summary here. The perturbative CEP amplitude, corresponding to the diagram shown in Fig.~\ref{fig:pCp}, can be written as
\begin{equation}\label{bt}
T=\pi^2 \int \frac{d^2 {\bf Q}_\perp\, \overline{\mathcal{M}}}{{\bf Q}_\perp^2 ({\bf Q}_\perp-{\bf p}_{1_\perp})^2({\bf Q}_\perp+{\bf p}_{2_\perp})^2}\,f_g(x_1,x_1', Q_1^2,\mu^2; t_1)f_g(x_2,x_2',Q_2^2,\mu^2;t_2)\; ,
\end{equation}
where  $Q_\perp$ is the transverse momentum in the central parton  loop (these partons are mainly gluons), with scale $Q_i^2 = Q_\perp^2$ in the forward proton limit (see e.g.~\cite{HarlandLang:2010ep} for the prescription away from this limit), and $\overline{\mathcal{M}}$ is the colour--averaged, normalised sub--amplitude for the $gg \to X$ process
\begin{equation}\label{Vnorm}
\overline{\mathcal{M}}\equiv \frac{2}{M_X^2}\frac{1}{N_C^2-1}\sum_{a,b}\delta^{ab}q_{1_\perp}^\mu q_{2_\perp}^\nu V_{\mu\nu}^{ab} \; .
\end{equation}
Here $a$ and $b$ are colour indices, $M_X$ is the central object mass, $V_{\mu\nu}^{ab}$ is the $gg \to X$ vertex, $q_{i_\perp}$ are the transverse momenta of the incoming gluons, and $t_i$ is the squared momentum transfer to the outgoing protons. The $f_g$'s in (\ref{bt}) are the skewed unintegrated gluon densities of the proton. These correspond to the distribution of gluons in transverse momentum $Q_i$, which are evolved  up to the hard scale $\mu_F$, such that they are accompanied by no additional radiation, as is essential for exclusive production. These unintegrated distributions can be obtained from the known integrated PDF
 using the prescription~\cite{K_WMR} based on the DGLAP evolution equation. 
We discuss this now.

\subsection{The necessity of embedding the Sudakov suppression   \label{sec:4.2}}

While the gluon momentum fractions $x_i$ are fixed by the mass and rapidity of the final state, the fractions ${x_i}'$ carried by the screening gluon must, in general, be integrated over at the amplitude level. However, for the dominant imaginary part of the amplitude $x' \ll x$, and it can be shown that the $f_g$'s may be simply written as\footnote{Unlike the usual unintegrated gluon distribution, a factor $\sqrt{T_g}$ occurs in (\ref{fgskew}) rather than $T_g$, since for $x'\ll x$ we have to account only for emission from the active gluon $x$.}
\begin{equation}\label{fgskew}
f_g(x,x',Q_\perp^2,\mu^2) \simeq \; \frac{\partial}{\partial \ln(Q_\perp^2)} \left[ H_g\left(\frac{x}{2},\frac{x}{2};Q_\perp^2\right) \sqrt{T_g(Q_\perp,\mu^2)} \right]\;,
\end{equation}
where $H_g$ is the generalised gluon PDF~\cite{Belitsky:2005qn}, which for CEP kinematics can be related to the conventional PDFs~\cite{Shuvaev:1999ce,Harland-Lang:2013xba}. In particular~\cite{Harland-Lang:2013xba}
\be
H_g\left(\frac x2,\frac x2,Q^2\right)=\frac{4x}{\pi}\int_{x/4}^1 dy\ y^{1/2}(1-y)^{1/2}g\left(\frac x{4y},Q^2\right).
\ee

 Note the occurrence of the Sudakov factor, $T_g$, in (\ref{fgskew}). It corresponds to the probability that no extra parton is emitted when two gluons fuse to make $X$. 
It is given by the resummation of the virtual loop corrections in the splitting functions $P(z)$ of the DGLAP evolution. Namely
\begin{equation}\label{tsp}
T_g(Q_\perp^2,\mu^2)={\rm exp} \bigg(-\int_{Q_\perp^2}^{\mu^2} \frac{{\rm d} k_\perp^2}{k_\perp^2}\frac{\alpha_s(k_\perp^2)}{2\pi} \int_{0}^{1-k_\perp/(\mu+k_\perp )} \bigg[ z P_{gg}(z) + n_F P_{qg}(z) \bigg]{\rm d}z \bigg)\;.
\end{equation}
where $\mu=M_X$~\cite{TF}.
It resums, to next--to--leading logarithmic accuracy, the virtual logarithms in $M_X^2/Q_\perp^2$ which occur when the loop momenta in virtual diagrams become soft and/or collinear to the external particle directions. That is, with these choices of the lower and upper cutoffs on the $k_\perp$ and $z$ integrals, it takes into account all terms of order $\alpha_s^n\ln^m(M_X^2/Q_\perp^2)$, where $m=2n,2n-1$. More physically, it corresponds to the (Poissonian) probability of no extra parton emission from a fusing gluon; that is, the probability that the gluon evolves from a scale $Q_\perp$ to the hard scale $\mu$ without additional real emission. This interpretation follows from the fact that these large higher--order logarithms are  being generated by a mismatch between the real and virtual corrections which  occurs due to the exclusivity requirement that no extra emissions be present. 

Note that if the Sudakov $T_g$ factor was neglected, then we would obtain an infrared divergent integral (\ref{bt}). In some earlier  papers
on Higgs-boson CEP, e.g. ~\cite{BL},\cite{Cudell:1995ki}~-~\cite{Boonekamp:2004nu}
it was assumed that at low $Q_\perp$ the cutoff of the integral (\ref{bt}) (or its analogue) 
is provided by the size of the proton or by confinement. Actually for a large mass, $M_X$, of the central system, the inclusion of
 $T_g$ makes the amplitude (\ref{bt}) convergent. The reason is that, at low $Q_\perp$, the value of $T_g(Q^2_\perp,\mu^2)$ decreases faster than any power of 
 $Q_\perp$. Indeed, in the leading double log approximation, $T_g$ is
 \be
 T_g(Q^2_\perp,\mu^2=M^2_X)=\exp\left(-\frac{N_c\alpha_s}{4\pi}\ln^2\left(\frac{Q^2_\perp}{M^2_X}\right)\right)\ .
 \ee

In summary, the potential infrared divergency of the CEP amplitude (\ref{bt}) in the absence of the $T_g$ factor, means that we cannot consider  the Sudakov suppression as some external gap survival probability, but must include it
in the calculation of the original QCD-induced CEP matrix element from the very beginning.

\begin{figure}
\begin{center}
\includegraphics[width=18cm]{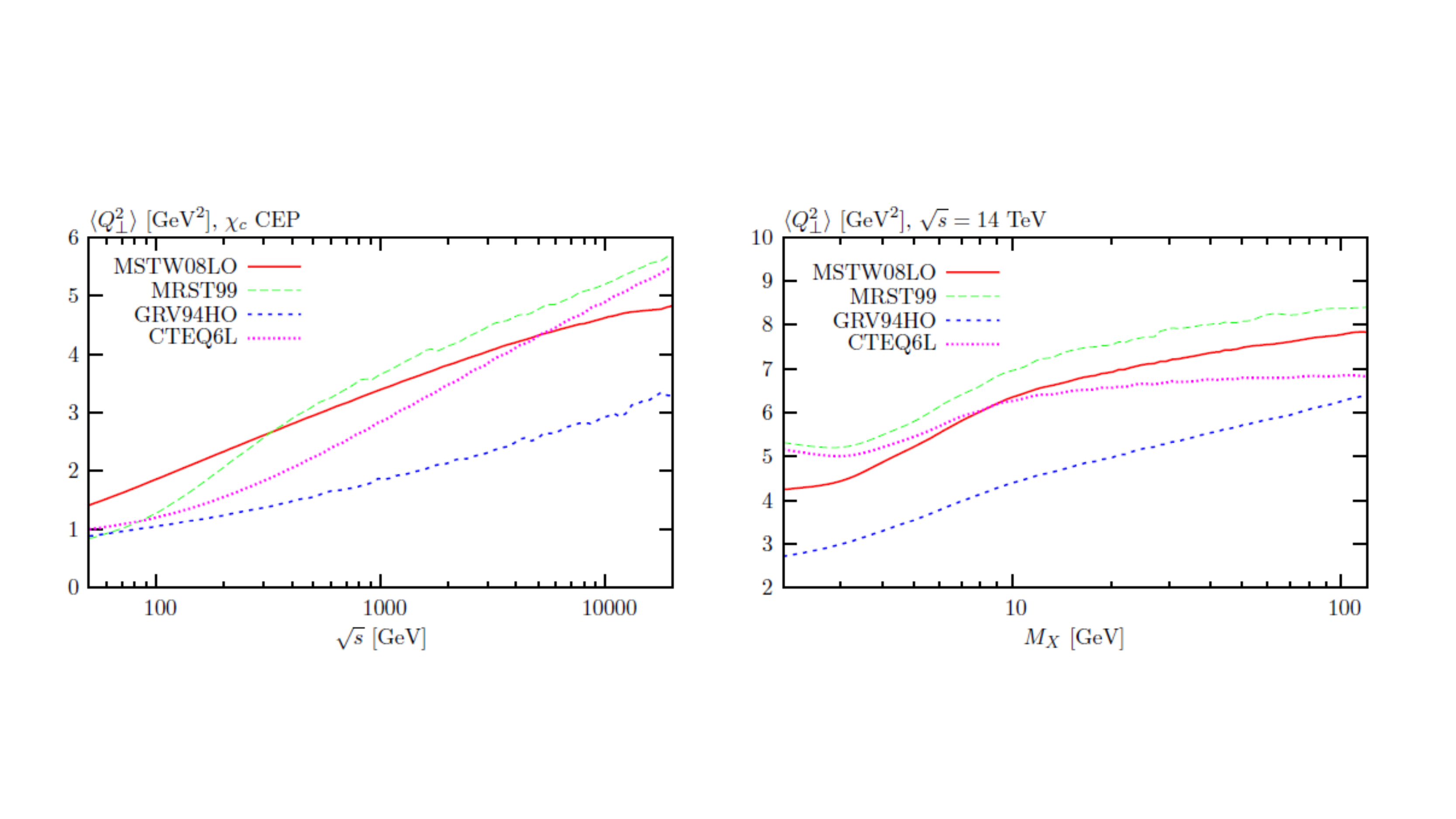}
\vspace{-2.5cm}
\caption{The average gluon squared--transverse momentum $\langle Q_\perp^2\rangle$ in the integrand of (\ref{bt}) for the production of a $0^+$ scalar particle in the forward proton ($p_\perp=0$) limit, as a function of the c.m. energy $\sqrt{s}$, and of the object mass $M_X$, for different choices of the gluon PDF (GRV94HO~\cite{Gluck94}, MSTW08LO~\cite{Martin:2009iq}, CTEQ6L~\cite{Pumplin:2002vw} and MRST99~\cite{Martin:1999ww}). The figure is taken from \cite{Harland-Lang:2014lxa}.}
\label{fgcomp1}
\end{center}
\end{figure}

Accounting for the Sudakov effect, that is selecting processes without any jet bremsstrahlung, we obtain a convergent integral with an  integrand which has  a saddle point (a maximum) at a relatively large 
$Q_\perp=Q_0$.
In the double log approximation the position of the saddle point is \cite{KMRh}
\be
\label{q0}
Q_0=M_X\exp\left(-\frac{\pi}{N_c\alpha_s(Q_0)}\right).
\ee
For LHC energies and $M_X\sim 100$ GeV,  the value of $Q_0^2=4-6$ GeV$^2$.
The relatively large value of $Q_0$ and the fact that $Q_0$ increases with $M_X$ (see (\ref{q0}) and Fig.~\ref{fgcomp1}) justifies the applicability of perturbative QCD for the main amplitude of the central exclusive production of a high-mass system.

\begin{figure}
\begin{center}

\includegraphics[width=15cm]{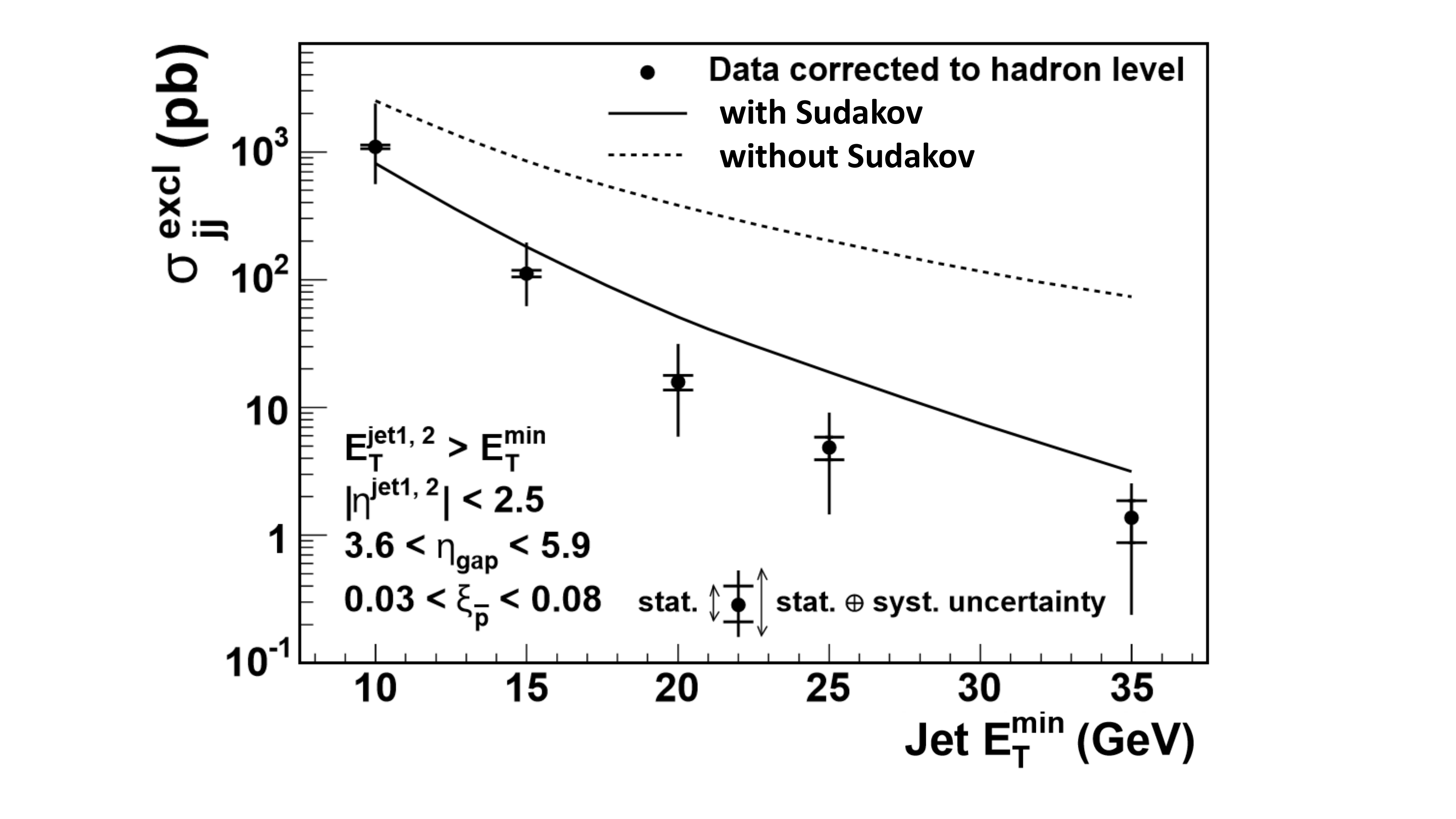}
\caption{The $E_T$ dependence of the exclusive dijet cross section. The predictions including (omitting) the Sudakov suppression are shown by continuous (dotted) curves. The figure is adapted from \cite{Aaltonen:2007hs}.
}
\label{S2ba}
\end{center}
\end{figure}

The inclusion of the $T_g$-factor results in a much steeper behaviour of the CEP cross section \cite{Aaltonen:2007hs}. For example, as demonstrated in Fig.~\ref{S2ba}, for exclusive dijet production at the Tevatron, the comparison of a calculation with the Sudakov effect included, to one without $T_g$, gives 5 times smaller cross section at jet energy $E_T=5$ GeV and up 25 times  smaller at $E_T=35$ GeV. The $E_T$ dependence obtained with the Sudakov suppression included  agrees well with the data; see also Fig.11 of~\cite{KMR}.

\subsection{Helicity amplitudes for CEP processes   \label{sec:4.3}}

It is interesting to decompose (\ref{Vnorm}) in terms of the on--shell helicity amplitudes for the central exclusive production of objects $X$. If we neglect small off--shell corrections of order $\sim {\bf q}_\perp^2/M_X^2$, and, for simplicity, omit the colour indices, then
\begin{align}
q_{1_\perp}^i q_{2_\perp}^j V_{ij} =\begin{cases} &-\frac{1}{2} ({\bf q}_{1_\perp}\cdot {\bf q}_{2_\perp})(T_{++}+T_{--})\;\;(J^P_z=0^+)\\ 
&-\frac{i}{2} |({\bf q}_{1_\perp}\times {\bf q}_{2_\perp})|(T_{++}-T_{--})\;\;(J^P_z=0^-)\\ 
&+\frac{1}{2}((q_{1_\perp}^x q_{2_\perp}^x-q_{1_\perp}^y q_{2_\perp}^y)+i(q_{1_\perp}^x q_{2_\perp}^y+q_{1_\perp}^y q_{2_\perp}^x))T_{-+}\;\;(J^P_z=+2^+)\\ 
&+\frac{1}{2}((q_{1_\perp}^x q_{2_\perp}^x-q_{1_\perp}^y q_{2_\perp}^y)-i(q_{1_\perp}^x q_{2_\perp}^y+q_{1_\perp}^y q_{2_\perp}^x))T_{+-}\;\;(J^P_z=-2^+)
\end{cases}
\label{Agen}
\end{align}
where the $J_z^P$ indicate the parity and spin projection on the $gg$ (beam) axis, and $T_{\lambda_1\lambda_2}$ are the corresponding $g(\lambda_1)g(\lambda_2)\to X$ helicity amplitudes,
 see~\cite{Harland-Lang:2014lxa,HarlandLang:2010ep} for more details. In the forward proton limit (i.e. with outgoing protons with $p_\perp=0$),  the only non-vanishing term in (\ref{Agen}), after the $Q_\perp$ integration (\ref{bt}), is the first one. This is the origin of the selection rule~\cite{Kaidalov:2003fw,Khoze:2000mw,Khoze:2000jm} which operates in this exclusive process: the $J^P_z=0^+$ quantum numbers for the centrally produced state are strongly favoured. More generally, away from the exact forward limit the non-$J^P_z=0^+$ terms in (\ref{Agen}) do not give completely vanishing contributions to the $Q_\perp$ integral. We find that 
\begin{equation}\label{simjz2}
\frac{|A(|J_z|=2)|^2}{|A(J_z=0)|^2} \sim \frac{\langle p_\perp^2 \rangle^2}{\langle Q_\perp^2\rangle^2}\;,
\end{equation}
 which is typically of order $\sim 1/50-1/100$, depending on such factors as the mass of the central object, c.m.s. energy $\sqrt{s}$ and the choice of the PDF set~\cite{Harland-Lang:2014lxa,HarlandLang:2010ep}. 
 The on--shell decomposition (\ref{Agen}) is used throughout, unless otherwise stated.

\subsection{Eikonal $S^2_{\rm eik}$ caused by soft multiple interactions   \label{sec:4.4}}

The expression (\ref{bt}) corresponds to the amplitude for the exclusive production of an object $X$ (that is, with no accompanying perturbative emission) in a short--distance interaction. However, as we are requiring that there are no other particles accompanying this final state, we must also include the probability that these are not produced in additional soft proton--proton interactions (or `rescatterings'), independent of the hard process, i.e. as a result of underlying event activity. This probability is encoded in the so--called `eikonal survival factor', $S^2_{\rm eik}$~\cite{Bjorken:1991xr,KMR,Ostapchenko:2010gt,Ryskin:2009tk},
\cite{Khoze:2006uj}~-~\cite{Gotsman:2014pwa}.

 The survival factor is conventionally written in terms of the proton opacity $\Omega(s,b)$. As was shown in subsection~\ref{sec:2.1} the proton opacity is related, via the usual elastic unitarity equations, to such hadronic observables as the elastic and total $pp$ cross sections.
 Thus, while the survival factor is a `soft' quantity which cannot be calculated using pQCD, it may be extracted from soft hadronic data~\cite{Khoze:2013dha,Khoze:2013jsa}.
  This survival suppression is found to be a sizeable effect, reducing the CEP cross section by about two orders of magnitude at LHC energies. 

Recall that the survival factor is not a simple multiplicative constant~\cite{HarlandLang:2010ep}, but rather depends on the distribution of the colliding protons in impact parameter space, or, in terms of momenta, it depends on the transverse momenta, $p_{\perp,1,2}$ of the outgoing protons.
In the simplest `one--channel' eikonal model, which ignores any internal structure of the proton, we can write the average suppression factor as
\begin{equation}\label{S2}
\langle S^2_{\rm eik} \rangle=\frac{\int {\rm d}^2 {\bf b}_1\,{\rm d}^2 {\bf b}_2\, |T(s,{\bf b}_1,{\bf b}_2)|^2\,{\rm exp}(-\Omega(s,b))}{\int {\rm d}^2\, {\bf b}_1{\rm d}^2 {\bf b}
_{2t}\, |T(s,{\bf b}_1,{\bf b}_2)|^2}\;,
\end{equation}
where ${\bf b}_{i}$ is the impact parameter vector of proton $i$, so that ${\bf b}={\bf b}_1+{\bf b}_2$ corresponds to the transverse separation between the colliding protons, with $b = |{\bf b}|$.  $T(s,{\bf b}_1,{\bf b}_2)$ is the CEP amplitude (\ref{bt}) in impact parameter space, and $\Omega(s,b)$ is the proton opacity discussed above; physically, $\exp(-\Omega(s,b))$ represents the probability that no inelastic scattering occurs at impact parameter $b$. 

While the rescattering probability  depends only  on the magnitude of the proton transverse separation $b$, the hard matrix element may have a more general dependence. More specifically, $T(s,{\bf b}_1,{\bf b}_2)$ is the Fourier conjugate of the CEP amplitude (\ref{bt}), i.e. we have
\begin{equation}\label{Mfor}
T(s,{\bf p}_{1_\perp},{\bf p}_{2_\perp})=\int {\rm d}^2{\bf b}_1\,{\rm d}^2{\bf b}_2\,e^{i{\bf p}_{1_\perp}\cdot {\bf b}_1}e^{-i{\bf p}_{2_\perp}\cdot {\bf b}_2}T(s,{\bf b}_1,{\bf b}_2)\;,
\end{equation}
where the minus sign in the ${\bf p}_{2_\perp}\cdot {\bf b}_2$ exponent is due to the fact that the impact parameter ${\bf b}$ is the Fourier conjugate to the momentum transfer ${\bf q}={\bf p}_{1_\perp}-{\bf p}_{2_\perp}$. We can therefore see that (\ref{S2}) is dependent on the distribution in the transverse momenta ${\bf p}_{i_\perp}$ of the scattered protons, being the Fourier conjugates of the proton impact parameters, ${\bf b}_i$. This connection can be made clearer by working instead in transverse momentum space, where we should calculate the CEP amplitude including rescattering effects, $T^{\rm rescatt}$. By integrating over the transverse momentum ${\bf k}_\perp$ carried round the Pomeron loop (represented by the grey ellipse\footnote{The ellipse is a simple graphical representation of the total eikonal effect which arises from all the different inelastic interactions of the incoming protons. Recall that the elastic amplitude which satisfies the unitarity equation, (\ref{un}), is expressed in terms of the same opacity $\Omega$ that describes the probability of inelastic interactions, $G_{\rm inel}$.} 
labeled `$S_{\rm eik}^2$' in Fig.~\ref{fig:pCp}) we obtain the amplitude which includes rescattering corrections in the following form
\begin{equation}\label{skt}
T^{\rm rescatt}(s,\mathbf{p}_{1_\perp},\mathbf{p}_{2_\perp}) = \frac{i}{s} \int\frac{{\rm d}^2 \mathbf {k}_\perp}{8\pi^2} \;T_{\rm el}(s,{\bf k}_\perp^2) \;T(s,\mathbf{p'}_{1_\perp},\mathbf{p'}_{2_\perp})\;,
\end{equation}
where $\mathbf{p'}_{1_\perp}=({\bf p}_{1_\perp}-{\bf k}_\perp)$ and $\mathbf{p'}_{2_\perp}=({\bf p}_{2_\perp}+{\bf k}_\perp)$, while $T^{\rm el}(s,{\bf k}_\perp^2)$ is the elastic $pp$ scattering amplitude in transverse momentum space, which is related to the proton opacity via
\begin{equation}\label{sTel}
T_{\rm el}(s,k^2_\perp)=2s \int {\rm d}^2 {\bf b} \,e^{i{\bf k_\perp} \cdot {\bf b}} \,T_{\rm el}(s,b)=2is \int {\rm d}^2 {\bf b} \,e^{i{\bf k_\perp} \cdot {\bf b}} \,\left(1-e^{-\Omega(s,b)/2}\right)\;.
\end{equation}
 We must add (\ref{skt}) to the `bare' amplitude excluding rescattering effects to give the full amplitude, which we can square to obtain the CEP cross section including eikonal survival effects
\begin{equation}\label{Tphys}
\frac{{\rm d}\sigma}{{\rm d}^2\mathbf{p}_{1_\perp} {\rm d}^2\mathbf{p}_{2_\perp}} \propto |T(s,\mathbf{p}_{1_\perp},\mathbf{p}_{2_\perp})+T^{\rm rescatt}(s,\mathbf{p}_{1_\perp},\mathbf{p}_{2_\perp})|^2 \;,
\end{equation}
where here (and above) we have omitted the dependence of the cross section on all other kinematic variables for simplicity.

In this way the expected soft suppression is given by 
\begin{equation}\label{seikav1}
\langle S_{\rm eik}^2\rangle= \frac{\int {\rm d}^2{\bf p}_{1_\perp}\,{\rm d}^2{\bf p}_{2_\perp}\,|T(s,\mathbf{p}_{1_\perp},\mathbf{p}_{2_\perp})+T^{\rm rescatt}(s,\mathbf{p}_{1_\perp},\mathbf{p}_{2_\perp})|^2}{\int {\rm d}^2{\bf p}_{1_\perp}\,{\rm d}^2{\bf p}_{2_\perp}\,|T(s,\mathbf{p}_{1_\perp},\mathbf{p}_{2_\perp})|^2}\;.
\end{equation}
It can readily be shown that (\ref{S2}) and (\ref{seikav1}) are equivalent. As we expect, the soft suppression factor depends on the proton transverse momenta, and so may have an important effect on the distribution of the outgoing proton ${\bf p}_{i\perp}$, via (\ref{Tphys}). A simplified approach, where the soft survival suppression is simply included in the CEP cross section as an overall constant factor will completely omit this effect. We also note that as the survival factor depends on the $p_\perp$ structure of the hard process, the average suppression will depend, as we will see later,  on the spin and parity of the object.

 Until now, we have considered the absorptive correction caused by diagrams with only elastic intermediate states. That is, we accounted only for the protons between the $S_{\rm eik}$ blob in Fig.~\ref{fig:pCp} and the main QCD amplitude.
However the contribution of heavier, $N^*_j$, states is not negligible.
 There may be the `quasi-elastic' $pp\to N^*_j p$ scattering accompanied by the sequential $N^*_j p\to p+X+p$  CEP amplitude. 

 It is convenient to describe such a processes in terms of the Good-Walker \cite{GW} eigenstates $\phi_i$ which at high energies `diagonalize' the matrix of the diffractive $N^*_j\to N^*_k$ transitions. In this formalism an arbitrary incoming state (including the proton) can be written as the superposition
 \be
 |j\rangle~=~\sum_k a_{jk}|\phi \rangle_k\ ,\;\;\;\; |p\rangle~=~\sum_k a_k|\phi_k\rangle
 \ee
 and the opacity $\Omega_{ik}$ corresponds to the {\em elastic} 
 $\phi_i+\phi_k\to \phi_i+\phi_k $ scattering. Since the $\phi_i$ are the eigenstates there are no quasi-elastic excitations such as $\phi_i+\phi_k\to \phi_j+\phi_k $.  

In this multi-channel approach the gap survival factor due to {\it eikonal} rescattering of the
Good-Walker eigenstates 
 at a fixed impact paramter ${\mathbf b}$, is
\begin{equation}  S^2_{\rm eik}({\mathbf b})~ = ~\frac{\left| {\displaystyle\sum_{i,k}} |a_{i}|^2~|a_{k}|^2~{\mathcal 
M}_{ik}({\mathbf b})~\exp(-\Omega_{ik}(s,{\mathbf b})/2)\right|^2}{\left|{\displaystyle \sum_{i,k}}
|a_{i}|^2~|a_{k}|^2~{\mathcal M}_{ik}({\mathbf b}) \right |^2} \,.
\label{eq:c3pp}
\end{equation}
Formally the main CEP amplitude ${\mathcal M}_{ik}$ can be calculated perturbatively (for a large mass $M_X$). The problem, however, is that we do not know how to re-distribute the partons given by the global analysis between the components $\phi_i$. One has to apply one or another hypothesis.
Since actually we are dealing with very low $x$ partons it looks natural
to assume that in each eigenstate $\phi_i$ the parton density is proportional to the value of the corresponding absorptive cross section $\sigma_i$. This approach is used in the majority of calculations.

Another, quite natural possibility, is to expect that the state of the smallest size contains mainly the valence quarks and a rather small number of `wee' (low $x$) gluons which are concentrated mainly in the large size state. That is, different states $\phi_i$ have a different $x$ distribution of its partons. This will lead to the dependence of $S^2_{\rm eik}$ factor on  mass and rapidity of the central system (see e.g.~\cite{Kaidalov:2001iz}). 

\begin{figure}
\begin{center}
\vspace{-7cm}
\includegraphics[width=18cm]{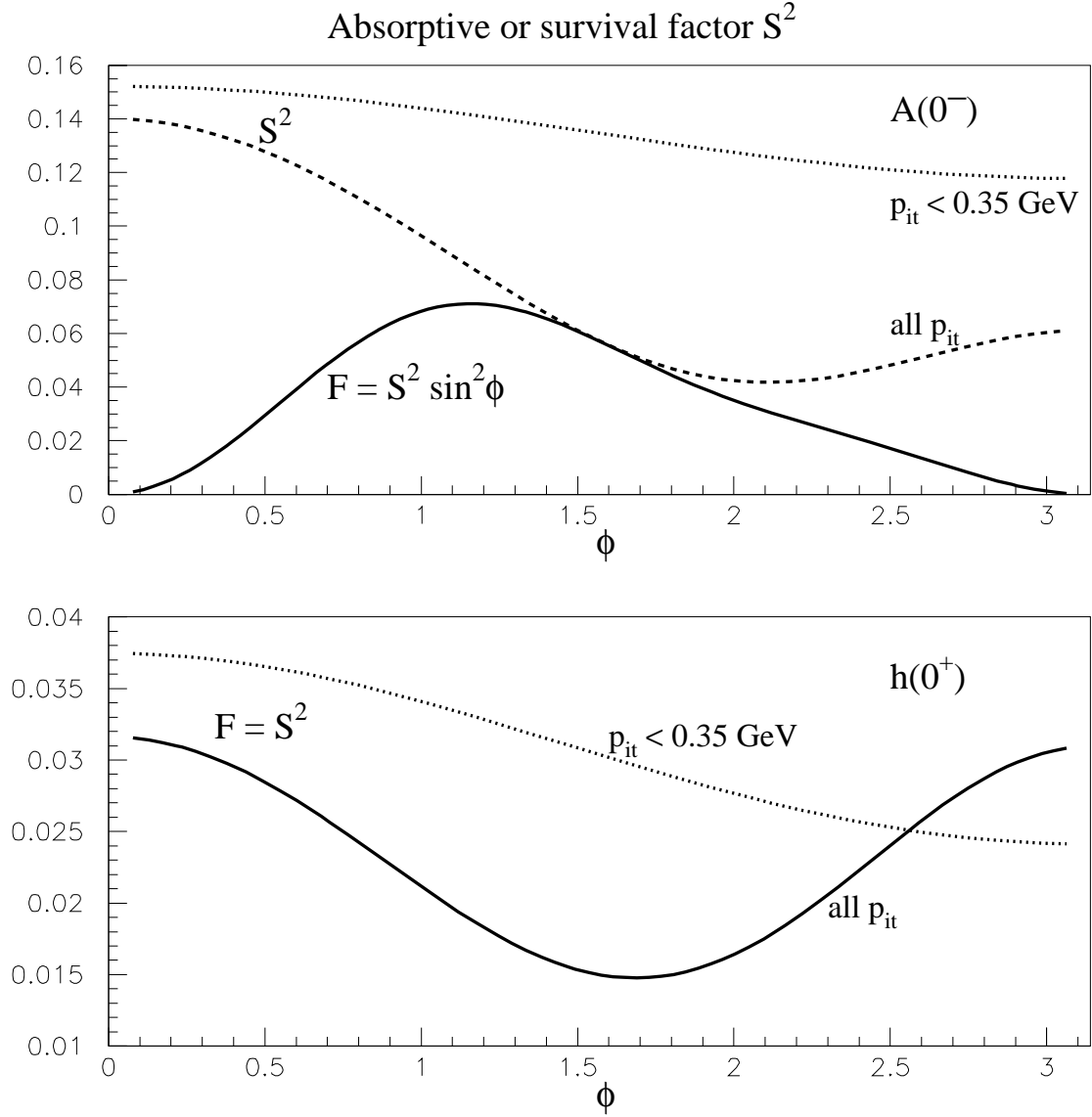}
\caption{The suppression factors $S_{\rm eik}^2$ of $h(0^+)$ and $A(0^-)\equiv h(0^-)$ Higgs production via the process $pp\to p + H + p$ at the LHC, arising
from rescattering effects. The outgoing protons are integrated 
over (i)~all $p_{i\perp}$ and (ii)~$p_{i\perp} < 0.35$ GeV (dotted curves). For illustration, the continuous curve for  $h(0^-)\equiv A(0^-)$ production includes the general  $\sin\phi$ behaviour of the bare amplitude. The figure is taken from \cite{Kaidalov:2003fw}.}
  \label{fig:5} 
  \end{center}
\end{figure}

\begin{figure}[t]
\begin{center}
\vspace{-2cm}
\includegraphics[width=18cm,trim=3cm 0 0 0]{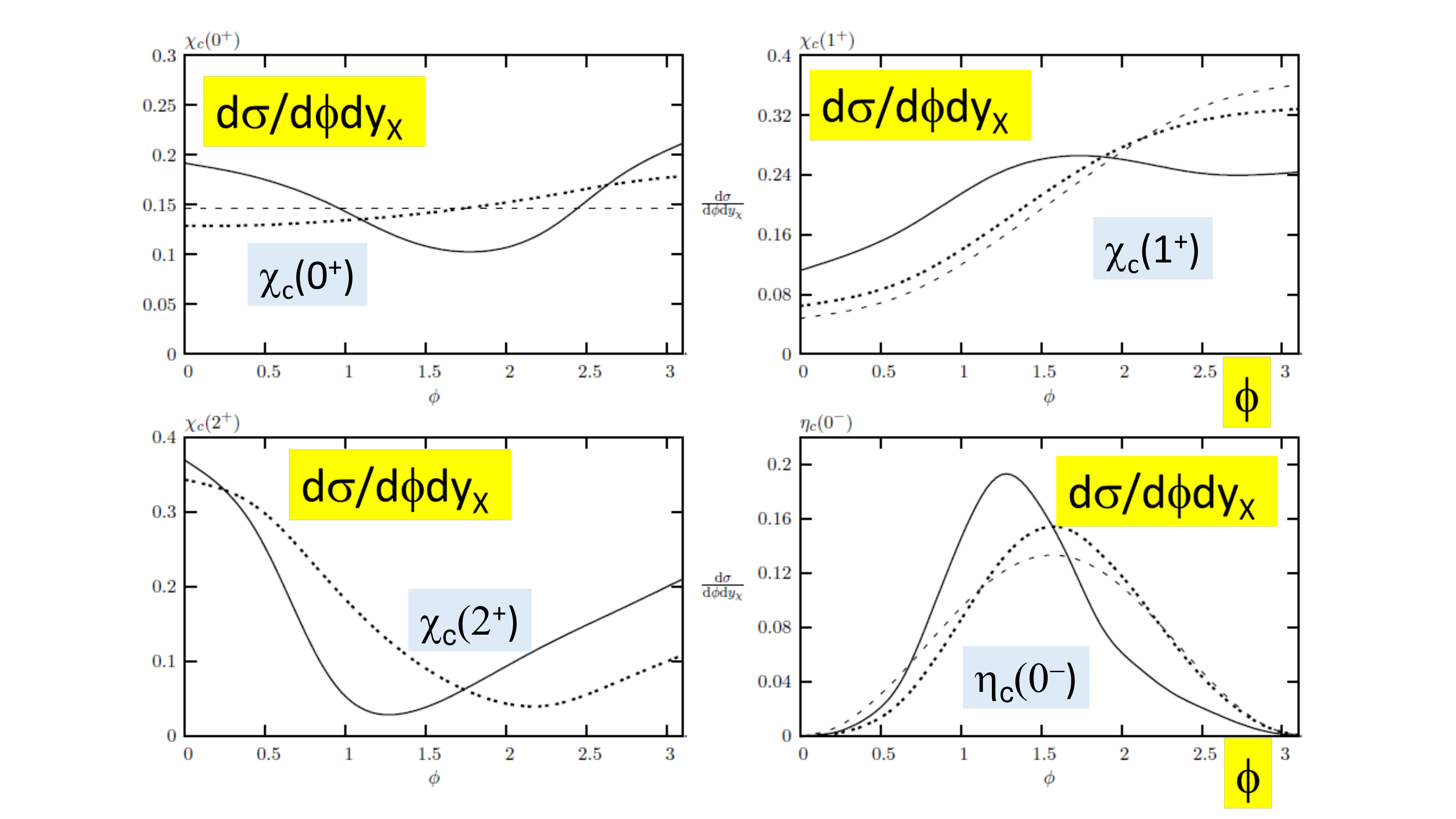}
\caption{Distributions (in arbitrary units) within the perturbative framework of the difference in azimuthal angle of the outgoing protons for the CEP of different $J^P$ $c\overline{c}$ states at $\sqrt{s}=14$ TeV and rapidity $y_X=0$. The solid (dotted) line shows the distribution including (excluding) the survival factor, calculated using the two--channel eikonal model~\cite{KMRsoft}, while the dashed line shows the distribution in the small $p_\perp$ limit, using the vertices of (\ref{Agen}) and excluding the survival factor. The figure is taken from \cite{Harland-Lang:2014lxa}.}\label{surv1}
\label{dist1}
\end{center}
\end{figure}

\begin{figure}
\begin{center}
\includegraphics[width=20cm]{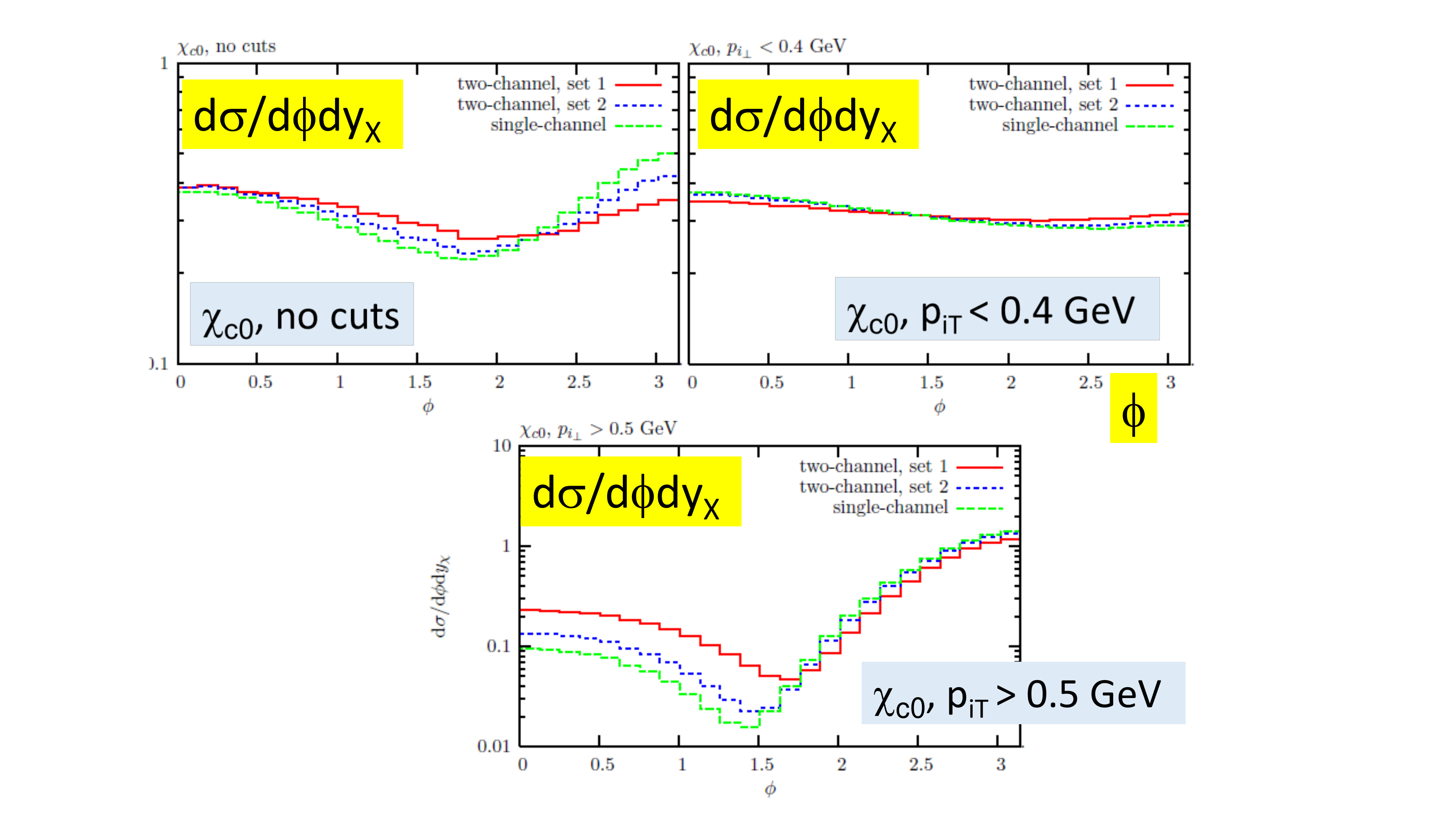}
\caption{Normalised distributions (in arbitrary units) of the difference in azimuthal angle between the outgoing protons for $\chi_{c0}$ CEP at $y_X=0$, with the survival factor $S^2_{\rm eik}$ calculated using the two--channel eikonal model~\cite{KMRsoft}, for two different choices of model parameters.  Also shown is the result of using the simplified single--channel eikonal approach~\cite{Khoze:2002nf}. Note that the `two--channel, set 1' $\chi_{c0}$ distributions are the same as those 
plotted in Fig.~\ref{dist1}. The figure is taken from \cite{Harland-Lang:2014lxa}.} \label{dist2}
\end{center}
\end{figure}

As seen from eqs.(\ref{Tphys},\ref{seikav1}) the $S_{\rm eik}$ suppression
{\em depends} on the matrix element of the CEP hard subprocess and on the transverse momenta, $p_{i_\perp}$, of the outgoing protons (see e.g.~\cite{Khoze:2002nf}). The effect can be seen in Figures~\ref{fig:5},\ref{surv1} and \ref{dist2}, where the dependences on the azimuthal angle between the protons momenta $\vec p_{1_\perp}$ and $\vec p_{2_\perp}$ are shown for the cases of heavy  
$0^+,\ 0^-$ scalars and for different $\chi_{cJ}$ charmonium states.


\begin{figure}
\begin{center}
\includegraphics[width=16.0cm]{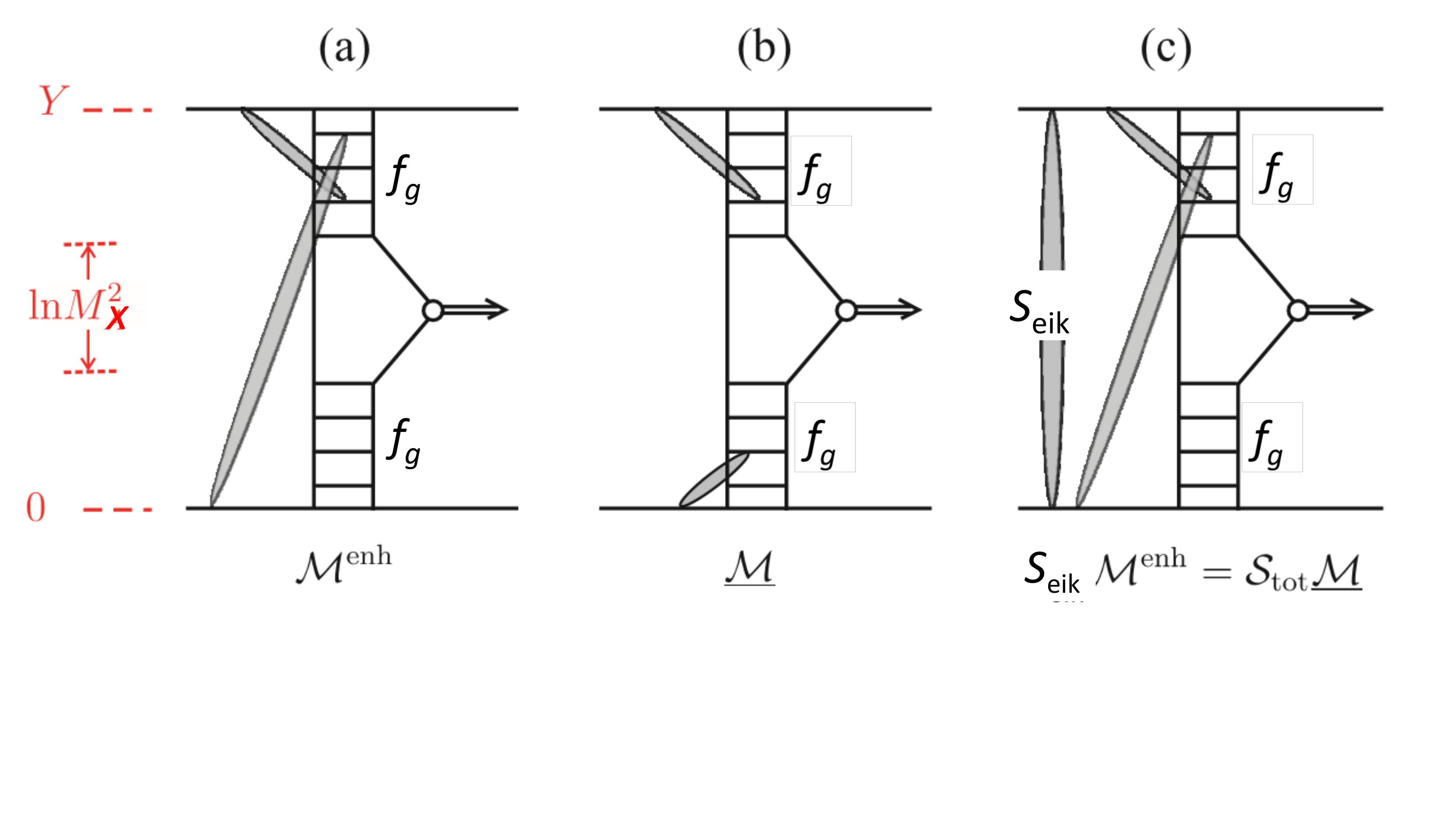}
\vspace{-2cm}
\caption[*]{(a): A symbolic representation of enhanced absorptive effects, (b): diagrams already included in $f_g$ extracted from global PDF analyses, and (c): diagrams relevant for the computation of $S^2_{\rm eik}S^2_{\rm enh}$.}
\label{fig:SS}
\end{center}
\end{figure}

\subsection{Enhanced screening \label{sec:4.5}}

Besides the effect of eikonal screening $S_{\rm eik}$, there is some suppression caused by the rescatterings of the protons with the intermediate partons~\cite{KMR,Ryskin:2009tk,Ryskin:2011qe} (inside the unintegrated gluon distribution $f_g$). This effect is described by the so-called enhanced Reggeon diagrams and usually denoted as $S^2_{\rm enh}$, see Fig.~\ref{fig:pCp} and Fig.~\ref{fig:SS}. The value of $S^2_{\rm enh}$ depends mainly on the transverse momentum of the corresponding partons, that is on the argument $Q^2_i$ in $f_g(x,x',Q^2_i,\mu^2)$ in (\ref{bt}), and depends only weakly on the $p_\perp$ of the outgoing protons~\cite{Ryskin:2011qe}. Working at LO (in collinear approximation) we have to
neglect such an effect. Due to the strong $k_t$-ordering the
transverse momenta of all the intermediate partons are very large
(i.e. the transverse size of the Pomeron is very small) and therefore the absorptive
effects are negligible. Nevertheless, this may not be true at  very low
$x$, where the parton densities become close to saturation and the
small value of the absorptive cross section is compensated by the large
value of the parton density.

 The absorptive corrections due to enhanced
screening must increase with energy. On the other hand,
 the gap survival factor $S^2_{\rm eik}$
already absorbs almost the whole contribution
from the centre of the disk. The parton essentially only survives eikonal
rescattering on the periphery, that is at large impact parameters $b$ where the parton density is rather small and the probability
of {\it enhanced} absorption is not large. This fact can be seen in
Ref. \cite{Watt}. There the momentum, $Q_s$, below which we may approach
saturation, was extracted from the HERA data in the framework of the
dipole model. Already at $b=0.6$ fm the value of $Q^2_s<0.3$ GeV$^2$
for the relevant $x\sim 10^{-6}$.

Note that the contribution of the diagrams Fig.~\ref{fig:SS}(b) is acutally already included  into the parton densities, $f_g$, given by the global parton analysis.
  The precise size of the $S^2_{\rm enh}$ effect is uncertain, but due to the relatively large transverse momentum (and so smaller absorptive cross section $\sigma^{\rm abs}$) of the intermediate patrons, it is only expected to reduce the corresponding CEP cross section by a factor of at most a `few', that is a much weaker suppression than in the case of the eikonal survival factor. The value of $S^2_{\rm enh}$ is also expected to depend crucially on the size of the available rapidity interval for rescattering $\propto \ln(s/M_X^2)$.

Combining these two effects, we may write down a  final expression for the CEP cross section at $X$ rapidity $y_X$
\begin{equation}\label{ampnew}
\frac{{\rm d}\sigma}{{\rm d} y_X}=\langle S^2_{\rm enh}\rangle\int{\rm d}^2\mathbf{p}_{1_\perp} {\rm d}^2\mathbf{p}_{2_\perp} \frac{|T(\mathbf{p}_{1_\perp},\mathbf{p}_{2_\perp})|^2}{16^2 \pi^5} S_{\rm eik}^2(\mathbf{p}_{1_\perp},\mathbf{p}_{2_\perp})\; ,
\end{equation}
where $T$ is given by (\ref{bt}) and the factor $\langle S^2_{\rm enh}\rangle$ corresponds to the enhanced survival factor averaged (i.e. integrated) over the gluon $Q_\perp$, while $S_{\rm eik}^2$ is simply given by
\begin{equation}
S_{\rm eik}^2(\mathbf{p}_{1_\perp},\mathbf{p}_{2_\perp})=\frac{|T(s,\mathbf{p}_{1_\perp},\mathbf{p}_{2_\perp})+T^{\rm rescatt}(s,\mathbf{p}_{1_\perp},\mathbf{p}_{2_\perp})|^2}{|T(s,\mathbf{p}_{1_\perp},\mathbf{p}_{2_\perp})|^2}\;,
\end{equation}
as can be seen by comparing (\ref{Tphys},\ref{seikav1}) and (\ref{ampnew}).

\begin{figure} [t]
\begin{center}
\vspace{-5cm}
\includegraphics[width=15cm]{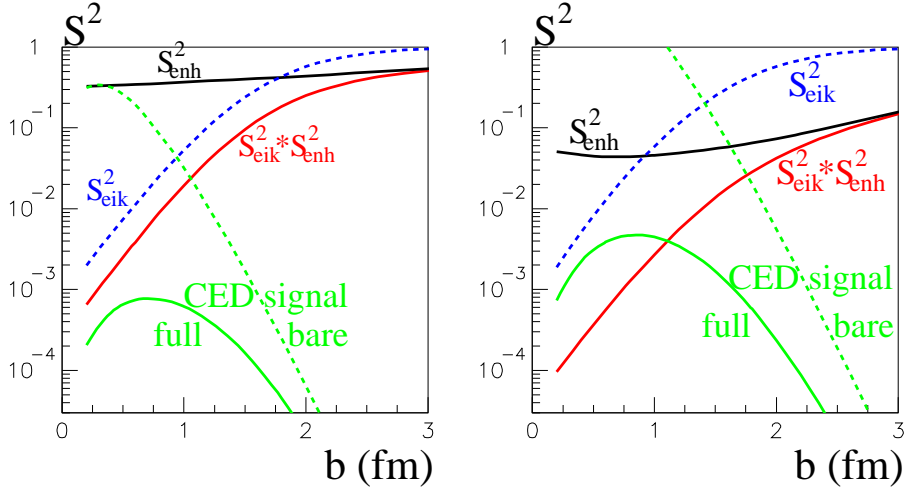}
\vspace{-6cm}\caption[*]{The survival factors for the Central Exclusive Diffractive  production (CEP) of a heavy $0^+$ system of mass $M_X$=150 GeV and $M_X$= 10 GeV. The figure is taken from \cite{Ryskin:2009tk}.}
\label{fig:SSb}
\end{center}
\end{figure}

The expected role of enhanced absorptive effects is shown in Fig.~\ref{fig:SSb} calculated in the
 model~\cite{Ryskin:2009tk,Ryskin:2009tj} where, to account for the $k_t$ dependence of the intermediate parton interactions, the pomeron was decomposed into the three different size (i.e. different $k_t$) components.\\
As is seen from Fig.~\ref{fig:SSb} the gap survival probability strongly depends on the value of the impact parameters typical for the 
`bare' CEP  amplitude. That is the value of $S^2$ is never universal.

The role of enhanced effects in diffractive dijet production was studied in~\cite{Ostap2}
 using the QGSJETII Monte Carlo~\cite{Ostapchenko:2013pia}.
It was shown  that, indeed, the gap survival suppression, 
caused by enhanced diagrams, $S^2_{\rm enh}$, is driven mainly by the value of impact parameter and only weakly depends  
on the initial energy, $\sqrt s$, and the transverse energy of the jets.

\section{Gap survival in soft processes   \label{sec:5}}

The eikonal survival factor $S^2_{\rm eik}$ for soft reactions with large rapidity gaps (LRG) is calculated in the same way as that in subsection~\ref{sec:4.4}. The only difference is that in the soft case the central matrix element occupies  
 a non-negligible area. Therefore the average impact parameter, $b$, becomes larger and correspondingly we get a somewhat larger  value of $S^2_{\rm eik}$. 

The situation for the enhanced part $S^2_{\rm enh}$ is more  complicated. This contribution should be calculated in terms of the Reggeon diagram technique
accounting for triple-Pomeron and/or other multi-Pomeron vertices.
 At moderate energies these absorptive corrections are relatively small due to the small phenomenological value of the triple(multi)-Pomeron vertex~
\cite{Kaidalov:1973tc}~-~\cite{Luna:2008pp}.

 However, at larger energies the role of  $S^2_{\rm enh}$  increases with $\ln s$, since the position of each vertex is integrated over the whole available rapidity interval.
 
 Besides this, we have to account for the possibility of an interaction between the leading nucleons and the central secondary hadrons. For example, for the soft CEP production of a pion pair ($pp\to p~+~\pi^+\pi^-~+~p$) we have to consider not only the additional proton-proton rescattering, but also the interaction with the outgoing pions as well (see the discussion in~\cite{Harland-Lang:2013dia}).

 The role of enhanced diagrams is also important for leading nucleon production. If we select events where the outgoing nucleon momentum fraction $x_L$ is close to 1, then we actually study a process with a LRG,
 $\delta y\simeq \ln(1/(1-x_L))$. Any additional interaction which populates this gap should simultaneously decrease the value of $x_L$.
 
 The precise description of the leading neutron cross sections at the LHC is interesting since at low neutron transverse momenta  pion exchange  gives the dominant contribution. Thus we have the possibility to measure the pion-proton total cross section at unprecedented high energies~\cite{KMR-ln,Petrov:2009wr,Ryutin:2016hyi}.
 Note that in this case an additional interaction which does not populate the gap does not lead to an absorptive correction. In particular, we should not include in the value of the corresponding $G_{\rm inel}$ the  diffraction dissociation of the target proton (assuming that the leading neutron is created by the beam).  For a detailed description see~\cite{KMR-ln}.
 Finally, we recall, that for not too large $x_L$ (starting from $x_L<0.75-0.8$), the neutron spectra are distorted by the `migration'. After  rescattering the value of the neutron's $x_L$ is decreased and its transverse momentum is changed.

 \section{LRG processes caused by $WW$ fusion} 
 
Another important class of the reactions with a LRG are processes caused by $WW$-fusion. For example, we have Higgs boson production ($WW\to H$)  
 or the $WW\to W^+W^-$ subprocess or the production of some new BSM object.
Here we do not deal with exclusive kinematics, since the initial protons are destroyed.   Instead, since  $W$ exchange describes a {\it colourless} spin 1 interaction, the best way to select such events is to require a LRG within the rapidity interval covered by the $W$ exchanges
 \cite{Bjorken:1991xr,Dokshitzer:1987nc,Bjorken:1992er,Khoze:2003af}.

Since the rapidity gap in the $WW$-fusion process could be populated by particles from multiple
interactions and from pile-up events, selecting events with {\em no hadronic activity} would give a gap survival $S^2$ suppression which is too strong. It was proposed~\cite{Dokshitzer:1991he}
to impose cuts to ensure relatively energetic particles or high $E_T>E_0$ jets~\cite{Zep} and to consider
 events with rather high $E_T$ jets at the edge of the LRG interval. 
  Indeed, due to a large mass $M_W$, the factorization scale corresponding to a parton which creates the $W$ is large ($\sim M_W$). That is, the parton, created together with $t$-channel $W$-boson, produces a high $E_T$ jet. Note that when, instead of $W$-exchange, we deal with the gluon induced reaction, then in normal QCD evolution such a jet should be accompanied by other high $E_T$ jets in the adjoining rapidity interval. Thus the requirement to have no additional mini-jets may be sufficient to suppress the QCD-induced reactions without decreasing the $WW$-fusion cross section too much.
  
  When we calculate the gap survival $S^2$ for jets produced with $E_T>E_0$ in the vetoed interval \cite{Zep}, we keep only the small
part of the inelastic
cross section in the opacity.
 Usually such a survival factor is calculated using a general purpose Monte Carlo (MC) generator
with a MI option. In this way we can account for the particular experimental cuts more precisely. In
particular, one has to take care about jets radiated near the point of $t$-channel $W$ formation.
The experimental cuts may veto part of this radiation leading to some additional Sudakov-like
suppression. A MC approach is the best option to account for such effect.

\section{Photon mediated processes  \label{sec:photo}}

 At present, there are no immediate plans to construct a high energy $\gamma \gamma$ collider.   However, proton colliders, in particular the LHC, can already serve as photon--photon colliders, see e.g. 
\cite{Albrow:2008pn,N.Cartiglia:2015gve}.
Indeed the LHC opens the way to observe photon collisions with high energies extending to the TeV
scale, well beyond the range probed by previous colliders.
This provides a unique opportunity to search for new objects (see for example \cite{Khoze:2001xm}~-~\cite{Khoze:2010ba},
\cite{Harland-Lang:2015cta}~-~\cite{Harland-Lang:2017cax}, \cite{Piotrzkowski:2000rx}~-~\cite{Khoze:2017igg}); in particular  those  which do not participate in the strong (QCD) interaction, and to study the  deviations from the SM.
The problem is that the inclusive cross sections of photon-induced reactions are rather low in comparison with those for QCD processes. 
Therefore we have to impose experimental cuts in order to suppress the QCD background.

\subsection{The photon PDF \label{sec:7.1}}
For pure {\it inclusive} processes the photon distribution function of the proton, $x\gamma^p(x,Q^2)$, is now known to good accuracy \cite{LUXqed} based on DIS data, measured mainly at HERA and JLab.  Another possibility to obtain $\gamma^p(x,Q^2)$ is to include the photon as a parton in the DGLAP evolution. Then at high scales the majority of photons will be radiated from quarks, while the input contribution, $\gamma^p(x,Q^2_0)$, contains two parts.
 An {\it elastic (or coherent}) component where the photon is radiated from the proton as a whole, $p\to p\gamma$. This is the dominant component and is well known in terms of the measured proton form factors; see, for example, \cite{MGRADM}.  Then we have the {\it inelastic (or incoherent)} component, $ep\to e\gamma X$, which may be calculated from low $Q^2$ DIS and resonant data on the structure functions $F_1$ and $F_2$.
In conclusion the photon PDF is completely determined from electron-proton scattering data. 

Usually the photon-induced events are selected by requiring a LRG. So we have to account for the gap survival factor $S^2$.
The problem is that the {\it different} components of the photon PDF, $x\gamma^p(x,\mu^2)$, are suppressed by  {\it different} $S^2$ factors. The inelastic component coming from evolution corresponds to rather large transverse momenta, $k_t$, that is to  small impact factors $b$. For this component the suppression is quite strong $(S^2\sim 0.01)$ (see Fig.~\ref{fig:SSb}). On the other hand, the low $k_t$ photon coming dominantly from the elastic component (the $p\to p+\gamma$  vertex) occupies the large $b$ region where $S^2$ is close to 1.
This means that we cannot multiply the precisely known inclusive PDF $x\gamma^p(x,\mu^2)$ by a common value of 
$\langle S^2\rangle $. We have to calculate the effective photon flux from the beginning based on the DGLAP evolution and accounting for the survival factor of each component.

\subsection{Photon-photon fusion processes at the LHC  \label{sec:7.2}}

As we have mentioned, photon-photon fusion processes at the LHC can be studied
either in  CEP processes with tagged forward protons or in events with LRG between the outgoing
proton remnants and the central system (in order to suppress the large
QCD-mediated backgrounds).
In the latter case the role of absorptive corrections is
especially crucial, see \cite{Harland-Lang:2016apc} for details.

Recall that, in comparison with QCD-induced exclusive production of an heavy object, the cross section of photon-photon fusion {\it exclusive} processes is not so small. First,
due to the strong Sudakov suppression, for the production of a heavy object the effective exclusive
`Pomeron-Pomeron' (that is the gluon-gluon) luminosity becomes comparable or even smaller (for $M_X>0.5$ TeV, see Fig.~2 of~\cite{Khoze:2001xm}) than the luminosity for photon-photon CEP processes. Second, while the QCD-induced CEP cross section is suppressed by more than 4-5 orders of magnitude in comparison with the inclusive channels, the exclusive photon-photon fusion cross section is suppressed by only one order of magnitude
 (see e.g.\cite{Harland-Lang:2016kog}).

 The first experimental results on the production of dilepton and $WW$ pairs (see \cite{Chatrchyan:2011ci}~-~\cite{Hollar:2017njv}) via  photon-photon fusion
and on light-by light scattering \cite{Aaboud:2017bwk}  at the LHC had    
 demonstrated the very promising potential of the LHC as a photon-photon collider.
Indeed, already some new bounds have been set on the  manifestations of BSM physics.
 Also, note that the dilepton CEP via the photon-photon mechanism could potentially  be used for
monitoring the LHC luminosity, see \cite{Khoze:2000db, Krasny:2006xg}.
 
\begin{figure} 
\begin{center}
\includegraphics[scale=1.0]{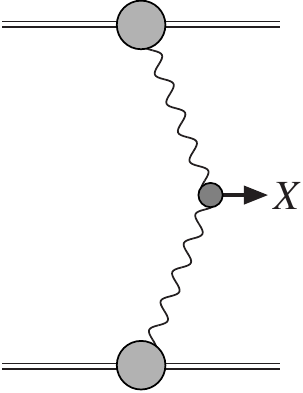}
\caption[*]{The photon--initiated CEP.}
\label{fig:CEP-phot}
\end{center}
\end{figure}
\begin{figure} 
\begin{center}
\includegraphics[height=3.5cm]{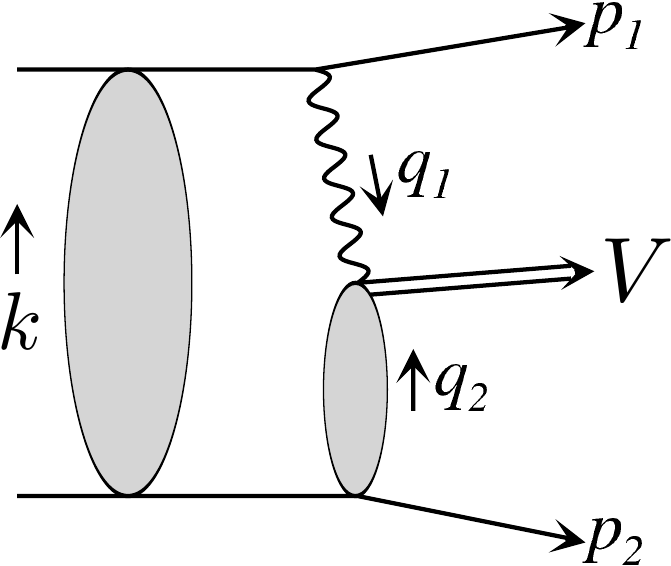}
\caption[*]{Schematic diagram for the exclusive photoproduction process $pp\to p+V+p$ with  screening corrections included.}
\label{fig:pVp}
\end{center}
\end{figure}

 Apart from the purely QCD--mediated interaction, 
(see Fig.~\ref{fig:pCp}) a CEP  event may be produced in
the collision of photons emitted from each proton (see Fig.~\ref{fig:CEP-phot})
or in the photoproduction process where one proton radiates the photon which then diffractively interacts with another proton
 as shown in Fig.~\ref{fig:pVp}. Indeed, data for the exclusive process $pp\to p+J/\psi +p$ \cite{LHCbJ} have been studied \cite{jones,jones2} to give information on the behaviour of the gluon PDF at low values of $x$. 

The necessary formalism for predicting the cross sections for these photon-induced CEP processes is given in the next subsection.
 For the first process, Fig.~\ref{fig:CEP-phot}, the QED initial state is particularly well understood, being simply given in terms of the known electromagnetic proton form factors
(see below), while  the impact of non--perturbative QCD effects is small. 
This allows a unique laboratory at the LHC  with which to observe QED mediated particle production, including  electromagnetically coupled BSM states.



\subsection{The basic formalism for photon-induced CEP processes \label{sec:photobare}}
In this subsection we introduce the formalism for predicting the rate of the exclusive processes shown in Figs.~\ref{fig:CEP-phot} and \ref{fig:pVp}. In general,
exclusive photon--exchange processes in $pp$ collisions are described in terms of the equivalent photon approximation~\cite{Budnev:1974de}. The quasi--real photons are emitted by the incoming protons $i=1,2$ with a flux given by
\begin{equation}\label{WWflux}
{\rm d}N^T(\xi_i)=\frac{\alpha}{\pi}\frac{{\rm d}q^2_{i_\perp} }{q_{i_\perp}^2+\xi_i^2 m_p^2}\frac{{\rm d}\xi_i}{\xi_i}\left(\frac{q_{i_\perp}^2}{q_{i_\perp}^2+\xi_i^2 m_p^2}(1-\xi_i)F_E(Q_i^2)+\frac{\xi_i^2}{2}F_M(Q_i^2)\right)\;,
\end{equation}
where $\xi_i$  and $q_{i_\perp}$ are the longitudinal momentum fraction and transverse momentum of the photon $i$, respectively.
  The functions $F_E$ and $F_M$ are given in terms of the proton electric and magnetic form factors, via
\begin{equation}
F_M(Q^2_i)=G_M^2(Q^2_i),\qquad F_E(Q^2_i)=\frac{4m_p^2 G_E^2(Q_i^2)+Q^2_i G_M^2(Q_i^2)}{4m_p^2+Q^2_i}\;,
\end{equation}
with
\begin{equation}
G_E^2(Q_i^2)=\frac{G_M^2(Q_i^2)}{7.78}=\frac{1}{\left(1+Q^2_i/0.71 {\rm GeV}^2\right)^4}\;
\end{equation}
in the dipole approximation, where $G_E$ and $G_M$ are the `Sachs' form factors. The modulus of the photon virtuality, $Q^2_i$, is given by
\begin{equation}\label{qi}
Q^2_i=\frac{q_{i_\perp}^2+\xi_i^2 m_p^2}{1-\xi_i}\;,
\end{equation}
 i.e. it is cut off at a kinematic minimum $Q^2_{i,{\rm min}}=\xi_i^2 m_p^2/(1-\xi_i)$.

We first give the formula for the cross section for the photoproduction of a state $V$, as shown in Fig.~\ref{fig:pVp}. If  the photon is emitted from proton $i$, then the result is simply given in terms of the photon flux (\ref{WWflux}) and the $\gamma p \to V p$ subprocess cross section
 \begin{equation}\label{sigphot}
 \sigma_{pp \to pVp}^i = \int {\rm d}N^T(\xi_i) \,\sigma_{\gamma p \to Vp}^i\;,
 \end{equation}
 integrated over the relevant phase space region. Note, however, that to be precise the flux (\ref{WWflux}) must be multiplied by a survival factor $S^2$ (see e.g. \cite{jones}). As the transverse momentum transferred by  photon exchange is typically much smaller than that due to the proton--Pomeron vertex, we may safely ignore interference effects, so that the total cross section is simply given by summing over $i=1,2$; that is, allowing for the emission of the photon from either  
 proton.
 
  For two--photon production $\gamma\gamma \to X$ of Fig.~\ref{fig:CEP-phot}, the corresponding cross section is
\begin{eqnarray}\label{siggam}
 \frac{{\rm d}\sigma_{pp\to pXp}}{{\rm d}\Omega}&=&\int\frac{{\rm
     d}\sigma_{\gamma\gamma\to X}(W_{\gamma\gamma})}{{\rm d}\Omega}\frac{{\rm
     d}L^{\gamma\gamma}}{{\rm d}W_{\gamma\gamma}}{\rm d W_{\gamma\gamma}}\;,
\end{eqnarray}
where $W_{\gamma\gamma}$ is the $\gamma\gamma$ c.m. energy. The $\gamma\gamma$ luminosity is  given by
\begin{equation}\label{lgam}
\frac{{\rm d}L^{\gamma\gamma}}{{\rm d}W_{\gamma\gamma}\,{\rm d}y_X}= \frac{2 W_{\gamma\gamma}}{s}\,n(x_1) \,n(x_2)\;,
\end{equation}
where $y_X$ is rapidity of $X$ and $x_{1,2}=(W_{\gamma\gamma}/\sqrt{s})\exp(\pm y_X)$, while $n(x_i)$ is the photon number density:
\begin{equation}
n(x_i)=\int  {\rm d}N^T(\xi_i) \,\delta(\xi_i - x_i)\;.
\end{equation}
Again the result must be corrected by including a gap survival factor.

\subsection{The inclusion of soft survival effects  \label{sec:photoscreen}}
For photon--mediated processes, survival effects can be included exactly as described in subsection~\ref{sec:4.4}.
However, some additional care is needed. From (\ref{Tphys}) we see that it is the {\it amplitude} for the production process that is relevant when including these effects. On the other hand, (\ref{sigphot}), (\ref{siggam}) and the flux (\ref{WWflux}) are defined at {\it cross section} level, with the squared amplitude for the photon--initiated subprocesses summed over the
 photon polarisations. 

To translate these expressions to the appropriate amplitude level, it is important to include the photon transverse momentum $q_\perp$ dependence in the appropriate way, corresponding to the correct treatment of the  photon polarisation, see~\cite{Khoze:2002dc}. To demonstrate this, we only consider the $F_E$ term in (\ref{WWflux}), but will comment on the contribution of the magnetic form factor, $F_M$, at the end. 

The schematic diagram for the  screened photoproduction amplitude is shown in Fig.~\ref{fig:pVp},  with the relevant momenta indicated. 
 Using the same decomposition that leads to (\ref{Agen}), the photoproduction amplitude 
behaves as 
\begin{equation}\label{tgam}
  T(q_{1_\perp}) \sim q_{1_\perp}^x (A^+ - A^-)+i q_{1_\perp}^y (A^+ + A^-)\;,
 \end{equation} 
  where $A^\pm$ is the $\gamma p\to V p$ amplitude for a photon of $\pm$ helicity. Recall that for a LRG  the photon polarization vector ${\vec e_i}$ is parallel to its transverse momentum $\vec q_{i_\perp}$ to good accuracy ($\sim O(\exp(-\Delta y)$, where $\Delta y$ is the size of rapidity gap). This is the origin of $q_{i_\perp}$ in (\ref{tgam}). If the photon is emitted from the other proton we simply interchange $1\leftrightarrow 2$. In the bare case ($q_{1\perp}=-p_{1\perp}$)  we just square (\ref{tgam}), and after performing the azimuthal angular integration, the cross terms $\sim p_{1_\perp}^x p_{1_\perp}^y$ vanish, leaving
  \begin{equation}\label{tgamsq}
  |T(p_{1_\perp}) |^2 \sim  p_{1_\perp}^2 \sigma_{\gamma p \to Vp}\;,
  \end{equation}
  where $\sigma_{\gamma p \to Vp}$ is the subprocess cross section summed over the incoming photon transverse polarisations. This is consistent with the $F_E$ term in (\ref{WWflux}) and with (\ref{sigphot}). Indeed a full treatment, keeping all prefactors and expressing the $pp \to pVp$ cross section in terms of (\ref{tgamsq}), leads to exactly these results, and is the essence of the equivalent photon approximation.
  
When calculating the screened amplitude it is crucial to correctly account for the explicit transverse momentum dependence as in (\ref{tgam}), with $q_{1_\perp} =k_\perp -{p'}_{1_\perp}$ included inside the integral (\ref{skt}); this vector structure of the amplitude can have a significant effect on the expected survival factor. More precisely for the photoproduction amplitude we take
\begin{equation}
T(q_{1_\perp},q_{2\perp}) = T' \frac{\sqrt{F_E(Q_i^2)}}{q_{1_\perp}^2+\xi_1^2 m_p^2}((q_{1_\perp}^x (A^+ - A^-)+i q_{1_\perp}^y (A^+ + A^-))\;,
\end{equation}
where $Q_i^2$ is given by (\ref{qi}). The factor $T'$ contains the transverse momentum dependence of the other proton,
 as well as the $\xi$ dependence (and other factors) in (\ref{WWflux}), and the $\gamma p$ ~c.m. energy $W_{\gamma p}$ dependence of the $\gamma p \to V p$ subprocess.  A 
 more detailed discussion of exclusive vector meson photoproduction is given in \cite{Harland-Lang:2015cta}.

For the two--photon initiated processes of Fig.~\ref{fig:CEP-phot}, the amplitude can be decomposed precisely as in (\ref{Agen}), i.e.
\begin{equation}
T(q_{1_\perp},q_{2\perp}) \sim -\frac{1}{2} ({\bf q}_{1_\perp}\cdot {\bf q}_{2_\perp})(T_{++}+T_{--})+\cdots\;,\label{Agenphot}
\end{equation}
where the $T_{\lambda_1\lambda_2}$ are now the $\gamma(\lambda_1)\gamma(\lambda_2)\to X$ helicity amplitudes, and we omit the overall factors for simplicity. Using this, it can readily be shown that the bare amplitude squared reduces to the correct expressions at the cross section level given in subsection~\ref{sec:photobare}, while with the inclusion of screening, it is again crucial to include this correct vector form of the amplitude; that is,  with the $q_{1_\perp}= k_\perp-{p'}_{1_\perp}$ and 
$q_{2_\perp}= -k_\perp-{p'}_{2_\perp}$ included inside the integral (\ref{skt}).

 Note that when including the absorptive correction, both photon transverse momenta depend on the common momentum $k_\perp$ transferred along the screening (loop) amplitude. This leads to a correlation between the two photon polarizations which affects the relative contributions of the  amplitudes $T_{\lambda_1\lambda_2}$.
In particular, for very small transverse momenta of recoil protons ($p_{i_\perp}\ll k_\perp$), the
screening correction acts mainly on the $T_{\pm\pm}$ amplitudes, whereas the correction  is suppressed by a factor $p_{i\perp}^2/k^2_\perp$ for the other  helicity states.
On the other hand, the $T_{\pm\pm}$ amplitudes vanish for non-forward massless dilepton production $\gamma\gamma\to l^+ l^-$. The massless dileptons are produced via the $T_{\pm\mp},\ T_{\mp\pm}$ helicity amplitudes. That is we
have {\em very} small absorptive corrections for  exclusive  $\mu^+\mu^-$ or $e^+e^-$ production (see the values of $S^2$  (last row) in 4th column of Table \ref{table:llww}). In principle, this fact has the potential to allow the use of such a process to measure the collider luminosity
(see \cite{Khoze:2000db} for details).

Note that we must also consider the contribution from the magnetic form factor $F_M$ in (\ref{WWflux}). While it is generally suppressed by $\xi^2$, for larger values of $M_X$ and/or production in the forward region, the corresponding value of $\xi$ may not be so small, and the contribution from this term may not be negligible.
This was discussed in more detail in~\cite{Harland-Lang:2015cta}.

Finally, recall that for central exclusive production induced by photon-photon fusion there
are {\em no} enhanced screening effects since the presence of LRG here is provided by elementary photon 
exchange and the corresponding interaction similar to that shown in Fig.~\ref{fig:SS}(a) is absent~\footnote{The corresponding 
diagram can only be introduced at the level of the ${\cal O}(\alpha_{\rm QED})$ corrections to the main process.}.

\subsection{$\gamma\gamma$ fusion reactions: comparing predictions with LHC data  \label{sec:2photo}}
\begin{figure} [t]
\begin{center}
\includegraphics[scale=0.7]{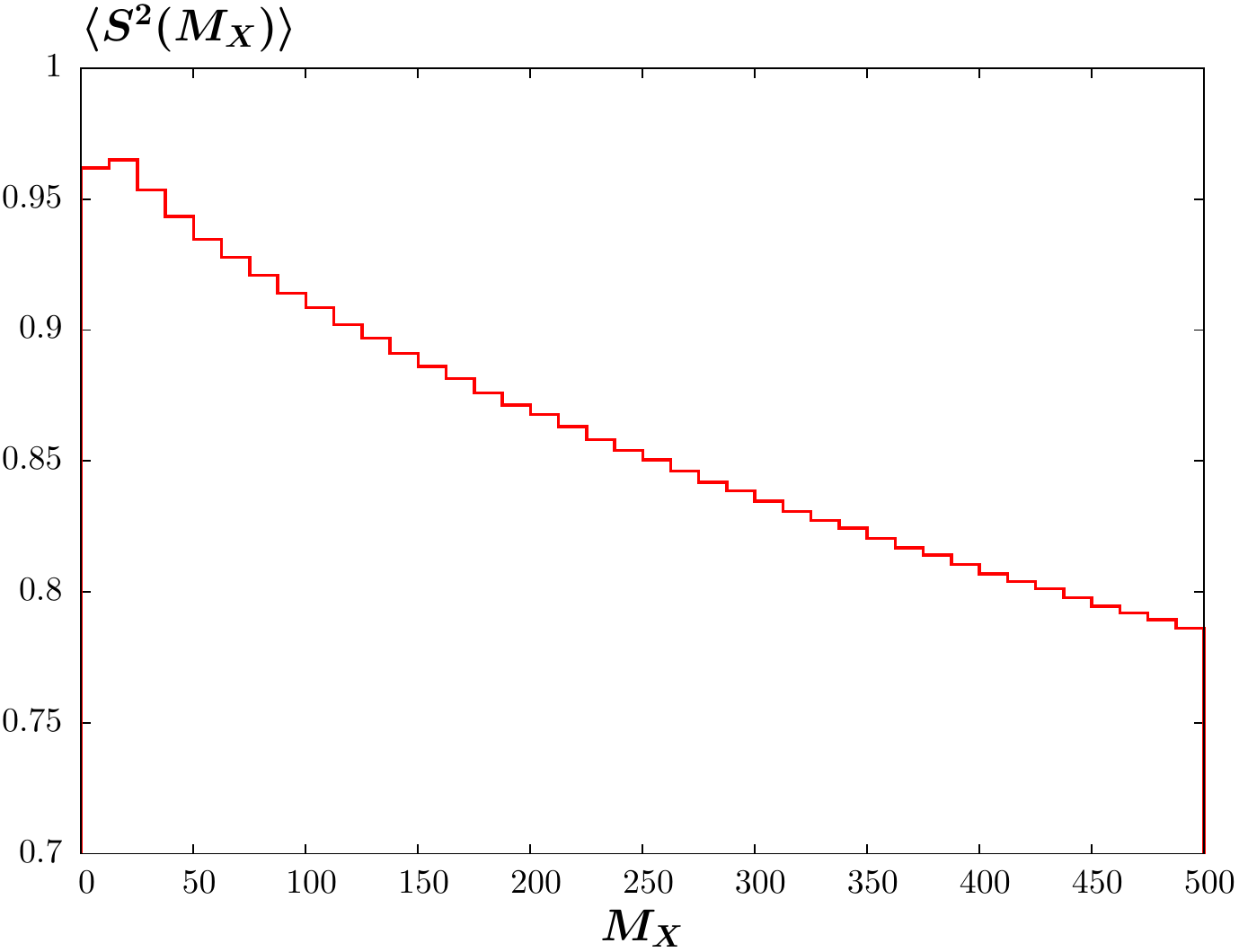}
\caption{Average survival factor $\langle S^2_{\rm elk} \rangle={\rm d}\sigma_{\rm scr.}/{\rm d}\sigma_{\rm bare}$ as a function of the central system invariant mass $M_X$ for lepton pair production, at $\sqrt{s}=14$ TeV. The leptons are required to have $p_\perp> 2.5$ GeV and $|\eta|<2.5$. The figure is taken from \cite{Harland-Lang:2015cta}.}
\label{fig:wwsurv}
\end{center}
\end{figure}

\begin{table}[h]
\begin{center}
\renewcommand\arraystretch{1.15}
\begin{tabular}{|c|c|c|c|c|}
\hline
&$\mu^+\mu^-$ &$\mu^+\mu^-$, $M_{\mu\mu}>2 M_W$&$\mu^+\mu^-$, $p_\perp^{\rm prot.}<0.1$ GeV&$W^+W^-$\\
\hline
$\sigma_{\rm bare} $& 6240 & 11.2& 3170& 87.5 \\
$\sigma_{\rm sc.} $ & 5990 & 9.58&3150 & 71.9 \\
$\langle S^2_{\rm eik}\rangle$ & 0.96 & 0.86&0.994 &0.82 \\
\hline
\end{tabular}
\caption{Cross section predictions (in fb) for exclusive muon and $W$ boson pair production at $\sqrt{s}=13$ TeV. The muons are required to have  $p_\perp> 5$ GeV and $|\eta|<2.5$, and are shown with and without an additional cut of $M_{\mu\mu}>2 M_W$, while in the $W$ boson case, no cuts are imposed. Results are shown for the `bare' and `screened' cross sections, i.e. excluding and including soft survival effects, respectively, and the resulting average suppression due to these is also given.}
\label{table:llww}
\end{center}
\end{table}
\noindent In this section, following \cite{Harland-Lang:2015cta},
we present a  brief selection of results for the two-photon exclusive production of lepton and $W$ boson pairs. As already mentioned these processes are now intensively studied
at the LHC, see \cite{Chatrchyan:2011ci}~-~\cite{Hollar:2017njv}. In Table~\ref{table:llww} we show predictions for the muon and $W$ boson pair production cross sections, with and without soft survival effects included. In the case of muon pair production we can see that, as expected from the previous discussion, 
 the average soft suppression factor is close to unity, due to the peripheral two--photon interaction, as well as the vanishing of the $T_{\pm \pm}$ amplitudes for massless leptons discussed in subsection~\ref{sec:photoscreen}. However,
as the invariant mass of the system increases, we expect the photon momentum fraction $x_\gamma \propto M_X$ to increase. This will lead to a higher average photon virtuality, see (\ref{qi}), and therefore for the average survival factor to be smaller for this less peripheral interaction. We also show the prediction for the same muon pair cross section, but subject to the requirement that $M_{\mu \mu}> 2 M_{W}$; while the suppression factor is still quite close to unity, it is clearly lower. This reduction in the survival factor with $M_X$ is seen more clearly in Fig.~\ref{fig:wwsurv},
where the average suppression is shown for lepton pair production as a function of the invariant mass of the pair; a very similar result is found for $W$ pair production. We also show in Table~\ref{table:llww} the total $W$ boson pair production cross section, where the suppression factor is smaller still, due to the different helicity structure of the production amplitudes (for which the $T_{\pm\pm}$ amplitudes are non--vanishing).  Finally, we show the muon pair production cross section, but with the outgoing protons required to have transverse momentum $p_\perp<0.1$ GeV. By imposing such a cut, the reaction is required to be highly peripheral. Moreover, for low $p_\perp$ the screening contribution to the dominant  $T_{\pm\mp},\ T_{\mp\pm}$ lepton amplitudes is suppressed by a factor $\langle p^2_\perp\rangle/\langle k^2_\perp\rangle$ (see subsection~\ref{sec:photoscreen}). Indeed,  it can be seen from Table~\ref{table:llww} that for $p_\perp <0.1$ GeV, the $S^2$ factor is extremely close to unity. On the other hand, as discussed in~\cite{HarlandLang:2012qz}, in the case when one or both protons dissociate the reaction is generally much less peripheral, and a proper inclusion of soft survival effects becomes crucial; this can lead to sizeable deviations in the data with respect to the result of, for example, the 
LPAIR MC~\cite{Vermaseren:1982cz,Baranov:1991yq}, 
which does not include these effects.

\begin{table}
\begin{center}
\renewcommand\arraystretch{1.15}
\begin{tabular}{|c|c|c|}
\hline
&$\mu^+\mu^-$ &$e^+e^-$\\
\hline
$\sigma_{\rm bare} $& 0.795 & 0.497 \\
$\sigma_{\rm sc.} $ & 0.742 & 0.459\\
$\langle S^2_{\rm eik}\rangle$ & 0.93 & 0.92\\
\hline
ATLAS data~\cite{Aad:2015bwa}&$0.628\pm 0.032\pm 0.021$&$0.428\pm 0.035\pm 0.018$\\
\hline
\end{tabular}
\caption{Cross section predictions (in pb) for exclusive muon and electron pair production at $\sqrt{s}=7$ TeV. The muons (electrons) are required to have  $p_\perp> 10(12)$ GeV, and in both cases $|\eta^l|<2.4$. Results are shown for the `bare' and `screened' cross sections, i.e. excluding and including soft survival effects, respectively, and the resulting average suppression due to these is also given. These are compared to ATLAS data~\cite{Aad:2015bwa}.}
\label{table:llATLAS}
\end{center}
\end{table}

The ATLAS collaboration have published a measurement of exclusive $\mu^+\mu^-$ and $e^+e^-$ production~\cite{Aad:2015bwa} in normal LHC running conditions, by vetoing on additional charged--particle tracks associated with the lepton vertex, and applying further corrections to extract the exclusive signal. This is compared to the MC predictions \cite{Harland-Lang:2015cta} in Table~\ref{table:llATLAS}. The bare cross sections are in both cases too high compared to the data, but a better agreement is achieved when survival effects are included. However, interestingly, while there is excellent agreement within uncertainties in the electron case, the prediction for the muon cross section lies $\sim$ 3 $\sigma$ above the data, i.e. a lower value of the average soft suppression appears to be preferred. 

 First of all the difference between the muon and the electron data (in comparison with the theoretical prediction) may be caused (leaving aside the violation 
 of $\mu~-~e$ universality ) by some purely experimental effects or this may be just statistical fluctuations.

 
 Next, it is not excluded that the cross section corresponding to proton dissociation, which was subtracted from the data, was overestimated. Recall that this cross section was calculated via the LPAIR MC program which does not account for the gap survival probability $S^2$. Besides this we note that, contrary to the bare photon exchange cross section which is proportional to 
 $1/p_{i_\perp}^2$, the screening (or absorptive) correction to the amplitude has {\em no} $1/p_{i_\perp}$ singularity. So selecting  the pure CEP photon initiated cross section based on its  $1/p^2_{i_\perp}$ behaviour (as was done) one may introduce an additional bias. (A detailed attempt is made in~\cite{Aad:2015bwa} to subtract this background and to account for any uncertainty on this in the systematic error on the data.) Further measurements, ideally differential in $M_{ll}$ and $p_\perp$ of the lepton pair, as well as with tagged protons, thus effectively eliminating the possibility of proton dissociation, will be of great use in clarifying this issue.   It would be the best to compare the differential $m_{ll}$ and 
 $p_\perp$  differential data distributions with those generated by \texttt{SuperChic}~~\cite{Harland-Lang:2015cta}. 


It is worth emphasising that as the two--photon production process is theoretically so well understood, that data for these processes represents a particularly clean probe of soft survival effects, when the outgoing protons are tagged.

These results highlight the importance of a proper treatment of screening corrections, which are 
often still not fully addressed in the literature.
It is worth emphasising that the impact parameter dependence of both the opacity $\Omega$ and the $\gamma\gamma \to X$ amplitude must be accounted for, and if this is omitted it will give misleading results. This is the case in for example~\cite{Dyndal:2014yea}, which has been compared to the recent ATLAS measurement~\cite{Aaboud:2017oiq} of exclusive muon pair production. The principle cause for the difference between these results and the \texttt{SuperChic} prediction~\cite{Harland-Lang:2015cta} is not the choice of model for the opacity $\Omega$ (which may have some genuine model variation) but rather the fact that the impact parameter dependence of the $\gamma\gamma\to \mu^+\mu^-$ amplitude is omitted in~\cite{Dyndal:2014yea}. This has been checked explicitly in~\cite{Harland-Lang:2015cta}.

\subsection{Light-by-light scattering: $\gamma\gamma\to\gamma\gamma$ in Pb-Pb LHC collisions \label{sec:lbyl}}

\begin{figure}[t]
\begin{center}
\includegraphics[height=7cm]{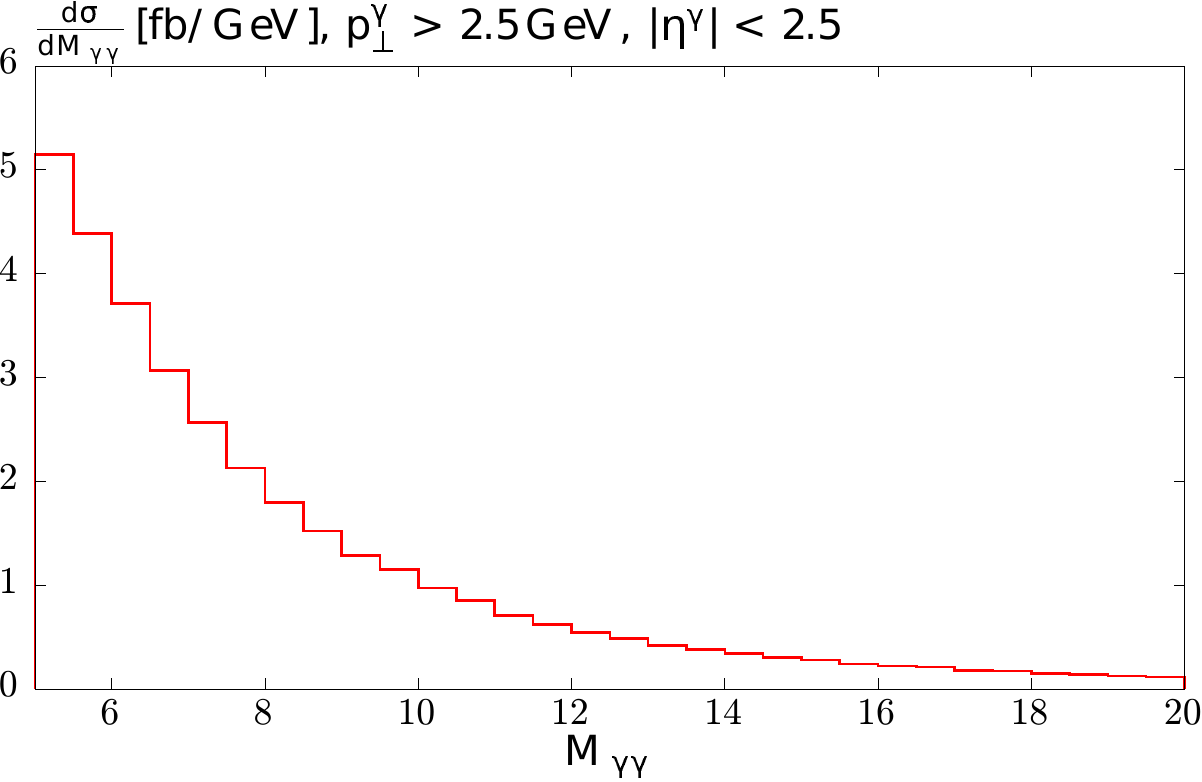}
\caption {The photon-photon invariant mass distribution for light--by--light scattering in $pp$ collisions at the 13 TeV LHC,
calculated using the \texttt{SuperChic} MC
\cite{Harland-Lang:2017cax}.}
\label{fig:s2m}
\end{center}
\end{figure}

It is interesting to consider  light--by--light scattering, $\gamma\gamma \to \gamma\gamma$, where, in the SM, the continuum process proceeds via an intermediate charged lepton, quark or $W$ boson box, see~\cite{dEnterria:2013zqi,Klusek-Gawenda:2016euz} for a detailed study. This process is  quite sensitive to BSM effects    
(see e.g.~\cite{Ellis:2017edi}~-~\cite{Knapen:2017ebd}), and
 nowadays it is particularly topical in view of the first direct evidence for this process by the ATLAS collaboration~\cite{Aaboud:2017bwk}, in Pb--Pb collisions. The predicted invariant mass distribution for 13 TeV  $pp$ collisions is shown in Fig.~\ref{fig:s2m}.

In addition to the light--by--light signal, it is  possible for the exclusive $\gamma\gamma$ final state to be produced via the QCD interaction $gg\to \gamma\gamma$ as in Fig.~\ref{fig:pCp}.
 In the ATLAS analysis, to estimate cross section for Pb-Pb collisions, the \texttt{SuperChic} prediction for $pp$ collisions was corrected by a factor of $A^2 R_g^2$, taken from~\cite{dEnterria:2013zqi}, where $A=208$ is the lead mass number and $R_g \approx 0.7$
 accounts for nuclear shadowing effects. In other words, up to the shadowing correction the predicted cross section in $pp$ collisions is simply scaled by the number of participating nucleons in the collision. However, this argument is not justified. In particular, due to the large inelastic nucleon-nucleon cross section, only those nucleons which are situated on the ion periphery (where the nucleon density is sufficiently small) may interact, while leaving the ions intact. A detailed calculation is therefore required, with the survival factor evaluated by correctly accounting for the geometry of the heavy ion collision. In this way, it was
found  that (for a large $A$) the CEP cross section in heavy ion collisions will instead scale like $\sim A^{1/3}$~\cite{Harland-Lang:2017cax}. We will therefore expect the QCD--initiated contribution to be lower than the simple $A^2$ scaling would suggest, although a precise numerical prediction is required to confirm the level of suppression.

\section{How to deal with LRG screening in $\gamma$-induced events   \label{rapgaps}}


As a rule, to select the relatively rare $\gamma\gamma$--initiated 
 events experimentally, some additional cuts must be imposed, including generally a requirement of a LRG on either side of the produced object. These cuts affect the incoming photon luminosity and require us to modify the corresponding photon distribution, which no longer corresponds to the usual inclusive one of subsection~\ref{sec:7.1}.
 
In earlier studies at the LHC, when no detection of forward protons was possible,
various selection cuts on
the final state in the central detectors were imposed in order to select a $\gamma\gamma$ enriched event sample. In such a case the interacting
protons may dissociate and
there will be an additional secondary particle production in some rapidity intervals.
However the events can still  have a diffractive topology,
and thus an attractive way to select photon exchange events is
to require a LRG between the centrally produced 
system ($W^+W^-$ or $l^+l^-$ pair, $J/\psi$ or $\Upsilon$, etc) 
and the forward outgoing secondaries. 
Such a strategy is mainly in operation in the recent LHC measurements.
Even when the AFP and CT-PPS forward proton spectrometers are engaged\footnote{For large luminosity of the collider, when 50 or more events may occur in each bunch crossing, precise timing of both protons may allow the position of the individual interaction of interest to be distinguished from other events. This should overcome the pile-up problem.},  in order 
to have an acceptance
for lower masses of the centrally produced system  we have to consider the configuration where the 
detection of only one proton is required while the other proton could dissociate into
an undetected system $p^*$, see \cite{CMS:2017uey,Hollar:2017njv}.  Recall that for double
tagging of the outgoing protons the spectrometer mass 
acceptance starts above about 200 GeV.

The potential for rapidity gap vetoes 
to select events with a diffractive topology in the low instantaneous luminosity runs
is in particular relevant at LHCb, 
for which the wide rapidity coverage allowed by the recently installed HERSCHEL forward detectors~\cite{Albrow:2014lta,McNulty}
 is highly favourable, while similar scintillation counters 
are also installed at ALICE~\cite{Schicker:2014wvk} and CMS~\cite{Albrow:2014lta,Albrow:2008az}.

\subsection{Factorization breaking effects in $\gamma\gamma$-induced processes \label{sec:8.1}}
Here we follow closely Ref.~\cite{Harland-Lang:2016apc} where the
$\gamma\gamma$--induced reactions are discussed for the case when LRG 
are present between the produced object and the outgoing proton dissociation products.
Provided the experimental rapidity veto region is large enough,
 the remaining contribution from non $\gamma\gamma$--initiated processes (e.g. standard Drell--Yan production) will be small, and can be suppressed with further cuts and subtracted using MC simulation, see for example~\cite{Aad:2015bwa,Chatrchyan:2013akv}.

When considering these processes, there are two important effects that must be correctly accounted for. First, the secondaries produced during the DGLAP evolution of the photon PDF may populate the LRG. 
This means that 
we cannot use the conventional inclusive PDF without additional restrictions.
A technique, is described in detail in \cite{Harland-Lang:2016apc}, which allows the construction of a new, modified, PDF where the evolution equation is supplemented by the condition that no $s$-channel partons are emitted in the LRG interval. 
Second, we have to account for the gap survival factor, $S^2$.
In this Section we focus on the inclusion of the $S^2$ suppression and illustrate the effects
this has on the $\gamma\gamma$ luminosity.

\begin{figure} [t]
\begin{center}
\includegraphics[scale=1.0,trim=12cm 12cm 10cm 12cm]{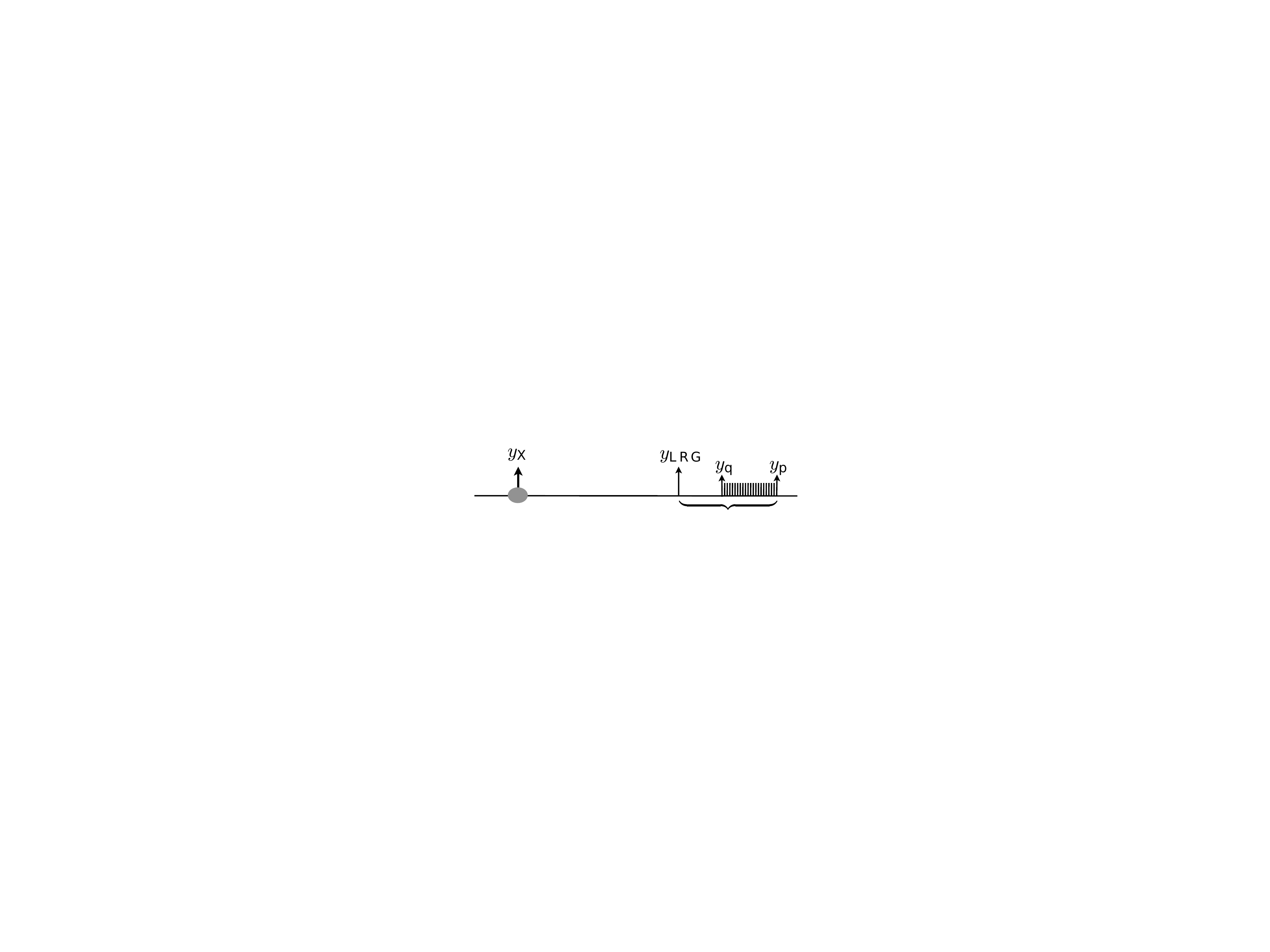}
\caption [*] {Schematic diagram corresponding to the diffractive topology described in text, where a quark of rapidity $y_q$ is emitted beyond the edge of the LRG region.}
\label{fig:gap}
\end{center}
\end{figure}

In Ref.~\cite{Harland-Lang:2016apc} it is shown how   to account for the
 rapidity gap veto in the $q\to q\gamma$ splitting associated with the LO DGLAP evolution of the photon PDF. 
In order to derive the modified photon PDF corresponding to a LRG, we require
that the quark, which radiates the photon,  be produced with rapidity greater than some $y_{\rm LRG}$,  corresponding to the end
of the experimentally defined gap. In this case
it is convenient to work in terms of the rapidity interval, $\delta=y_p-y_{\rm LRG}$ between the edge of the gap and the outgoing proton, in which the quark may be emitted, see Fig.~\ref{fig:gap}. The condition $y_q >y_{\rm LRG}$ in this notation takes the form
\be
y_p-y_q < \delta \ .
\label{dif}
\ee

 Recall, from subsection~\ref{sec:7.1},
 that the impact of survival effects depends sensitively on the subprocess, via the specific proton impact parameter dependence. Thus, for exclusive photon exchange processes, the low virtuality  (and hence transverse momentum) of quasi-real photon exchange, which corresponds to relatively large impact parameters between the colliding protons, leads to an average survival factor that is close to unity, while for the less peripheral QCD--induced exclusive processes the suppression is much larger.

\begin{figure}
\begin{center}

\includegraphics[height=5cm]{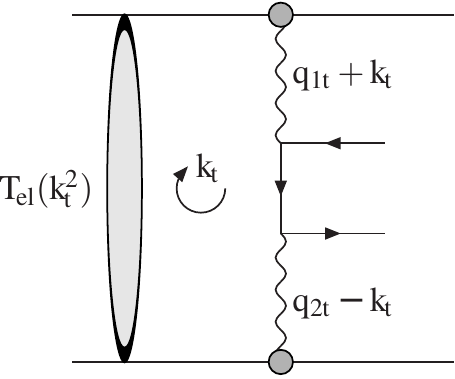}
\caption{Feynman diagrams for the screened amplitude for coherent photon--induced lepton pair production.}
\label{fig:survcoh}
\end{center}
\end{figure}

We noted that the photon PDF $x\gamma^p(x)$ has three main components: 
\begin{itemize}
\item[{(a)}]  photons emitted coherently by all the valence quarks via the `elastic' $p\to p+\gamma$ vertex, 
\item[{(b)}] `incoherent' emission with excitation of the proton, $p\to N^*+\gamma$ or $p\to Y+\gamma$ with a rather small mass $M_Y$,
\item[{(c)}] photons from DGLAP evolution (accounting for the LRG condition (\ref{dif})).
\end{itemize}
In case (a) the photon's transverse momentum is small and the gap survival probability is close to 1. In case (b) we have a  larger $q_t$ and a smaller $S^2$, while the contribution from DGLAP evolution (c) is strongly suppressed by $S^2$ factor.   

The numerical results for the average survival factors were calculated in
Ref.~\cite{Harland-Lang:2016apc}, using
model 4 of~\cite{Khoze:2149}, which applies a two--channel eikonal approach, in which the incoming proton is considered to be a coherent superposition of two diffractive Good--Walker eigenstates~\cite{GW}, 
each of which may scatter elastically.
The results for all combinations of photon PDF components from each proton are given in Table~\ref{table:surv}. A large range of expected suppression factors is evident. As anticipated, $S^2$ for the lower scale (and hence more peripheral) coherent production process being higher than for the higher scale evolution component. The survival factor for the incoherent component of the input PDF is seen to be particularly small.

\begin{table}
\begin{center}
\begin{tabular}{|c|c|c|}
\hline
$\langle S^2\rangle$& $M_X^2=200 \,{\rm GeV}^2$ & $M_X^2=10^4 \,{\rm GeV}^2$  \\ \hline
(coh., coh.)& 0.95 & 0.89 \\
(coh., incoh.) &0.84  & 0.76\\
(incoh., incoh.)  &0.18  & 0.18 \\
(evol., coh.) &0.83  & 0.74 \\
(evol., incoh.) &0.16  & 0.16\\
(evol., evol.) &0.097  &0.097 \\
\hline
\end{tabular}
\caption{Average survival factor for different components of the photon PDFs $\gamma^p(x,M_X^2)$, for different masses $M_X$ of the central system produced with rapidity $y_X=0$. The coherent, incoherent and evolution components are shown for (proton 1, proton 2).}
\label{table:surv}
\end{center}
\end{table}

These results have important implications for the standard factorisation formula
\begin{equation}
\label{fact}
\sigma(X)=\int {\rm d}x_1{\rm d}x_2 \,\gamma^p(x_1,\mu^2)\gamma^p(x_2,\mu^2)\,\hat{\sigma}(\gamma\gamma\to X)\;,
\end{equation} 
which implies that the photon flux associated with each proton $i$ can be factorised in terms on an independent PDF $\gamma^p(x_i,\mu^2)$.
Instead, this now depends on the state of the other interacting proton, through the influence this has on the survival factor. 
Physically, this is to be expected, as the survival factor is generated by additional soft proton--proton interactions, which then prevents all of the physics associated with the initial--state photon produced by a given proton being considered independently from the photon emitted from the other proton.
Analogous factorisation breaking effects have already been seen in the example of 
diffractive dijet production at the Tevatron, see subsection~\ref{sec:1.2}. There the predictions using the so--called diffractive PDFs were found to dramatically overshoot the data when naively applied to hadron--hadron collisions~\cite{Affolder:2000vb}. 
It is therefore not possible to discuss the impact of survival effects on the individual photon PDF. Instead, we have to consider  the $\gamma\gamma$ luminosity given by each particular component.

\subsection{The dependence on the rapidity gap: the choice of $\delta$ }

\begin{figure}
\begin{center}
\includegraphics[height=5.2cm]{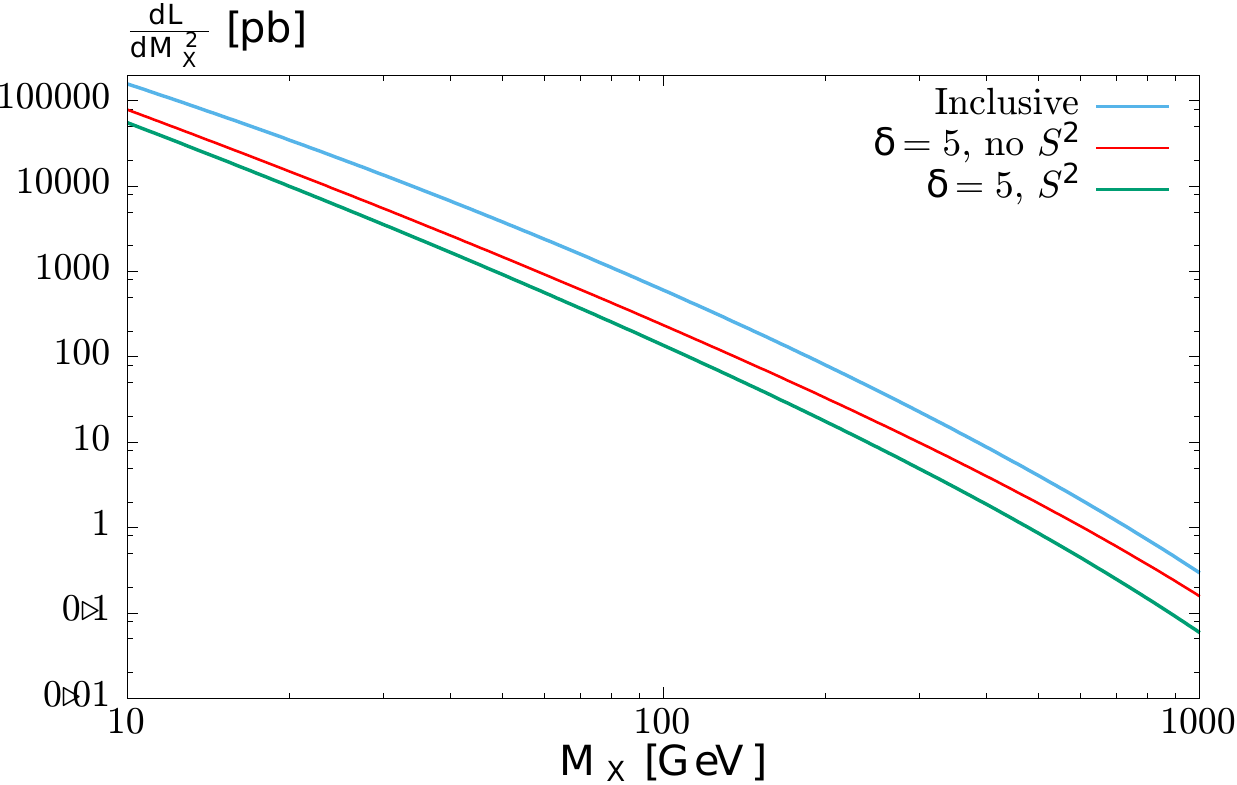}
\includegraphics[height=5.2cm]{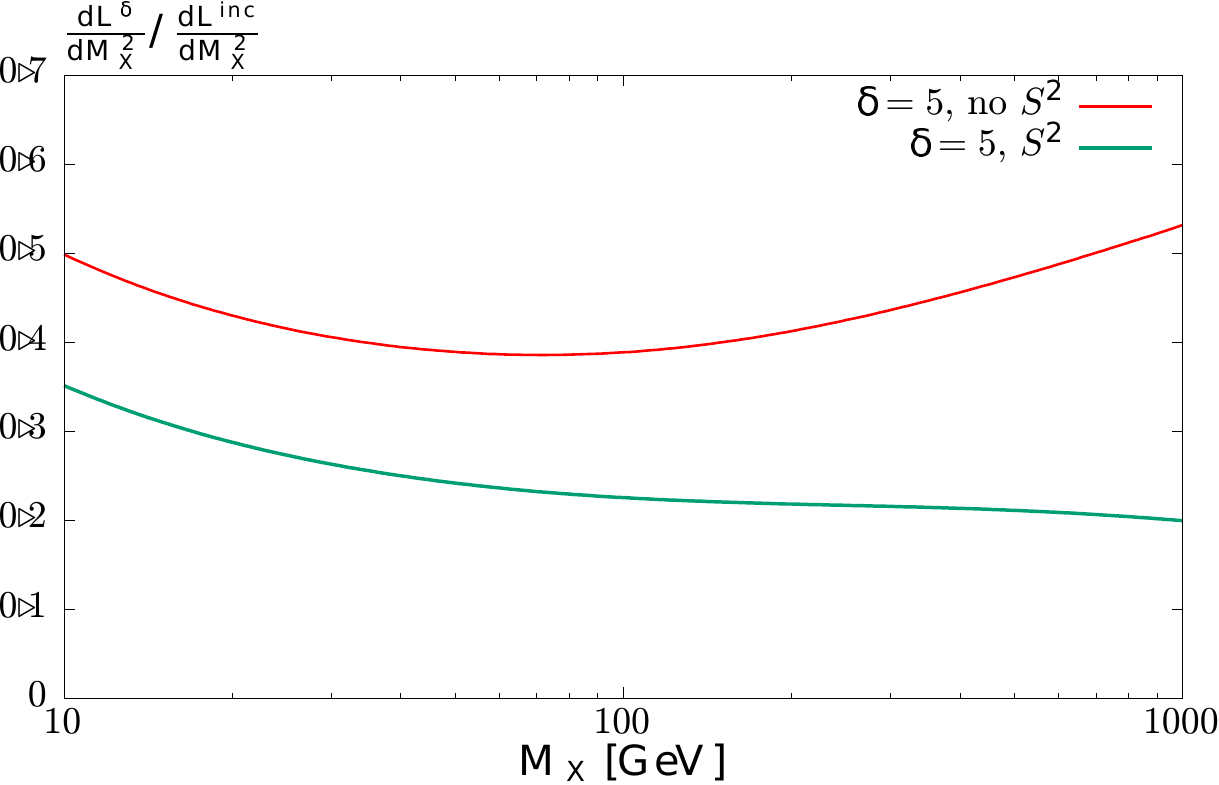}
\caption{$\gamma\gamma$ luminosity at $\sqrt{s}=13$ TeV in the inclusive and semi--exclusive cases, with $\delta=5$ for both protons defining the rapidity interval via (\ref{dif}). For demonstration purposes, the semi--exclusive luminosities are shown both with and without survival effects included. We show in the left-hand plot the absolute luminosities, and in the right-hand plot the ratios to the inclusive luminosity.}
\label{fig:lumi}
\end{center}
\end{figure}

 The photon--photon luminosity at $\sqrt{s}=13$ TeV is shown in the left plot of Fig.~\ref{fig:lumi}, for inclusive production and for semi--exclusive production (with $\delta=5$ for both protons defining the rapidity interval via (\ref{dif})). For illustration we show semi-inclusive production both with and without survival effects included. We can see that the inclusion of condition (\ref{dif}) leads to a  factor of $\sim 2$ reduction in the luminosity, 
while the inclusion of survival effects leads to a further suppression of  a similar size. That the suppression due to both effects is similar in size is not necessarily to be expected, and indeed for different choices of $\delta$ and/or $\sqrt{s}$, the relative contribution of these effects will differ. It is also interesting to consider how the suppression varies with the central system mass, $M_X$. This is shown on the plot on the right-hand side of Fig.~\ref{fig:lumi}. In both cases the dependence on $M_X$  is seen to be relatively mild. The suppression due to introducing the $\delta$ cut decreases at both low and high $M_X$, due to counteracting effects.
Increasing $M_X$ leads to a generally larger suppression due to the higher scale at which the PDF is evaluated, this also leads to a larger average $x$ value probed, for which the suppression is less, with similar, but opposite, effects for decreasing $M_X$. Once soft survival effects are included, however, the overall trend is simply a decrease with increasing $M_X$.

\begin{table}
\begin{center}
\begin{tabular}{|c|c|c|c|}
\hline
 & $\sigma^{\mu^+\mu^-}$, $p_\perp^\mu>10$ GeV &$\sigma^{\mu^+\mu^-}$,  $p_\perp^\mu>20$ GeV & $\sigma^{W^+W^- \to l^+ \nu\,l^-\overline{\nu}}$  \\ \hline
$\sigma^{\rm inc}$ [pb]& 12.2 & 2.4 &0.015\\
$\sigma^{\delta=3}/\sigma^{\rm inc.}$& 0.18& 0.16 &0.14\\
$\sigma^{\delta=7}/\sigma^{\rm inc.}$& 0.39& 0.36&0.31 \\
\hline
\end{tabular}
\caption{Cross section predictions for photon--induced muon and $W$ boson pair production at the $\sqrt{s}=13$ TeV LHC. The inclusive and ratio of semi--exclusive to inclusive cross sections are shown, with two choices of the emission region $\delta$. For the muon pair we show results for two values of cut on the muon transverse momenta. For the $W$ boson, the leptonic decay ($l=e,\mu$) is considered, with lepton transverse momenta $p_\perp^l>25$ GeV, and  missing transverse energy $E_\perp^{\rm miss}>20$ GeV. In all cases the lepton pseudorapidity is required to satisfy  $|\eta|^\mu<2.5$.}
\label{table:cross}
\end{center}
\end{table}

 Following ~\cite{Harland-Lang:2016apc}, for illustration purposes, we show in Table~\ref{table:cross}  the cross section predictions at the LHC for
  muon and $W$ boson pair production at $\sqrt{s}=13$ TeV
(see Table~\ref{table:cross}),
although the approach described above may be readily applied to other two--photon induced processes\footnote{In principle, it is also  possible to apply this approach to the semi-exclusive photoproduction of, for example, vector meson ($J/\psi$, $\Upsilon$...) states, see
~\cite{Harland-Lang:2016apc}.}.
Table~\ref{table:cross} shows the inclusive cross section and ratio of semi--exclusive to inclusive cases, with survival effects included and for two values of $\delta$ (applied to both proton sides). These choices are motivated by the experimental situation at the LHC. Considering just the central tracking detector of the ATLAS and CMS collaborations, we have an uninstrumented region beyond pseudo-rapidity  $|\eta|\sim 2.5$, which for $\sqrt{s}=13$ TeV corresponds to roughly $\delta\approx 7$. On the other hand, a much larger rapidity gap may be vetoed on using forward shower counters, installed at LHCb (the HERSCHEL detectors~\cite{Albrow:2014lta,McNulty}), 
CMS~\cite{Albrow:2008az} 
and ALICE~\cite{Schicker:2014wvk}. Roughly speaking, these extend the rapidity coverage out to $|\eta| \lesssim$ 8, which correspond to a representative value of $\delta\approx 2$. These may be considered as lower and upper bounds on experimentally realistic values of $\delta$. For other specific experimental configurations, the appropriate value may lie somewhere in between. 

However, $\delta=2$ represents a relatively small region for non--vetoed emission, which may be sensitive in particular to fluctuations due to fragmentation and hadronisation, see e.g.~\cite{Khoze:2010by}, and for which the transverse momentum of the outgoing quark due to DGLAP evolution satisfying (\ref{dif}) may not be sufficiently high that simple factorisation 
 holds. For example, if we take a characteristic $z\sim 0.2\, x$ in (\ref{dif}) we have $q_t \lesssim 1.5$ GeV. As well as potentially spoiling this factorisation, this relatively low scale indicates that such a veto may be sensitive to the incoherent input component of the photon PDF, for which the recoil quarks may be produced with sufficiently large $q_t$ to fill the rapidity gap. We thus present results for a somewhat higher value of $\delta=3$, in order to avoid too great a sensitivity to these effects.

We can see from Table~\ref{table:cross} that even for the relatively large $\delta=7$ there is expected to be a significant reduction, by a factor of $\sim 3$, in the predicted cross section relative to the inclusive case. For $\delta=3$ this reduces by a further factor of $\sim 2$, depending on the process. The greater 
suppression with increasing $M_X$, already seen in Fig.~\ref{fig:lumi}, 
is also evident.

\section{Gap survival, $S^2$, in Monte Carlo generators  \label{sec:9}}
At first sight, it seems that the best way to account for Large Rapidity Gaps (LRG), and other experimental cuts, is to use a general purpose MC. There are problems however.
\begin{itemize}

\item The cross sections of CEP or LRG processes are small. That is the efficiency of a general purpose MC will be extremely low. The probability to generate a low multiplicity exclusive event (or an event with a LRG) is less than $10^{-5} - 10^{-4}$. 

\item As a rule, a general purpose MC does not include the possibility of `soft screening' (like those shown schematically by the ellipses in Figs.~\ref{fig:1ab} and \ref{fig:pCp}) of the colour flow along the LRG.\footnote{An exception is the `SCI' (Soft Colour Interaction) MC~\cite{Edin:1995gi}. However even in this case,
instead of the full Generalized Parton Distribution function of eqs. (\ref{bt}) and (\ref{fgskew}), in the SCI model actually simplified one gluon exchange is implemented.}

\item The Multiple Interaction (MI) option as a rule does not account for the possibility of proton excitations discussed in 
subsection~\ref{sec:4.4}. That is, just a one-channel eikonal model is used.\footnote{The exception is the QGSJETII MC~\cite{Ostapchenko:2013pia} where the two-channel eikonal 
is implemented.}
\end{itemize}

Therefore few MC generators for CEP and/or LRG processes were written which produce only  the particles actually formed in the final state. The sub-amplitudes corresponding to the exchange across the LRG and the Sudakov-like 
suppression are calculated (in some sense) analytically. Here we have to mention 
the  RAPGAP MC for the electro(photo)-production process \cite{Jung:1993gf} and POMWIG~\cite{Cox:2000jt} where the 
 exchange across the LRG is described, not in terms of QCD, but  just by  
phenomenological Pomeron exchange. To be more precise, these generators use so-called diffractive PDFs (dPDF) measured at HERA in events with LRGs. These dPDF include the `soft' Pomeron flux multiplied by the parton distribution inside the Pomeron.
The FPMC generator~\cite{Boonekamp:2011ky} may operate either using the dPDFs or 
the Durham model described in subsection~\ref{sec:4.1}. Also the Durham model was implemented in ExHume~\cite{Monk:2005ji}. In all these generators the gap survival probability was not actually calculated. At the end, the cross section was multiplied by some mean number $\langle S^2 \rangle$.

The only two MCs where the gap survival factor, $S^2_{\rm eik}$, is {\em calculated} for each individual process and is implemented `differentially', so that we can see the role of $S^2$ in the transverse momenta distributions of the outgoing protons and other secondaries, is the \texttt{SuperChic} MC~\cite{Harland-Lang:2015cta} based on the Durham QCD model, and DIME~\cite{Harland-Lang:2013dia} used
for the CEP soft process (such as $pp\to p+\pi^+\pi^-+p$).

In the cases where, instead of a complete veto on the hadronic activity within the LRG region, we have just a veto on mini-jets with $E_T>E_0$ or a veto on  high $p_T$ hadrons, it is  better to use a general 
purpose MC like Pythia, Herwig or Sherpa, with the gap survival factor included via the MI option. An interesting example is the 
generator of exclusive final states~\cite{Lonnblad:2016hun} based on PYTHIA8  
where the Sudakov suppression is implemented in the most straightforward and precise way.  Then the effects of additional gluon radiation can be studied in detail. 
If we use a general purpose MC to evaluate the role of the $S^2$ factor, then we have to check how well the MI option is implemented and how well
the parameters used in MI option describe the elastic $pp$-cross section measured at the LHC.

Recall that at the moment, the best MC to calculate the gap survival probability to an additional soft interaction is QGSJETII \cite{Ostapchenko:2013pia}. It is the only MC which accounts for enhanced screening effects in addition to the eikonal 
$S^2_{\rm eik}$ suppression.

A more detailed review of the different MC generators for exclusive events is given in Chapter 2 of~\cite{N.Cartiglia:2015gve}.

\section{Cross checks on the gap survival calculations}

With the advent of forward proton detectors at the LHC, processes with LRGs can be very informative. However, each process has its own gap survival factor. They are not universal.
Moreover, to calculate the gap survival suppression we have to know the impact parameter, $b$, and $\sqrt{s}$
dependence of the proton opacity, $\Omega(b,s)$, and to account for the possible presence of different G-W eigenstates.

Therefore, first of all we have to check that the model used in the $S^2$ computations reasonably reproduces (i) the
$t$-dependence of the elastic
cross section at the given energy and (ii) the value of small-mass diffractive dissociation,
$\sigma^{\rm SD}_{{\rm low}M}$. In other words, it is crucial to have data on $d\sigma_{\rm el}/dt$ and on the low-mass
proton dissociation.\footnote{Note that there are accurate data for $d\sigma_{\rm el}/dt$, but very few measurements of $\sigma^{\rm SD}_{{\rm low}M}$.}

After these initial requirements are fulfilled, there are several processes that are able to check the quality of the model. For instance we may compare the predicted azimuthal correlation
between the transverse momenta of outgoing protons with that observed experimentally in the CEP dijet
process~\cite{Khoze:2002nf}, or in exclusive vector meson ($J/\psi$ or $\Upsilon$) production~\cite{Khoze:2002dc},
or in exclusive  $\chi_c$ production (see e.g. Fig.~8 and~\cite{Harland-Lang:2014lxa}).
We may also check the predictions for $W$ ($Z$) boson production with one LRG as was described
in~\cite{Early}.

The role of Sudakov suppression can be probed by studying the CEP three-jet events~\cite{Early} while the rapidity
dependence of the diffractive production of a heavy system (with one LRG) may be used to check the role of enhanced
survival factor~\cite{Ostap2,Early}.

\section{Conclusions}
Exclusive events, or the events with a Large Rapidity Gap
(LRG), are very informative, since they allow a study of rare processes
in a clean experimental environment. On the other hand the corresponding 
cross sections are strongly suppressed by the fact that the secondary particles 
from additional soft interactions, or from extra gluon radiation, may 
fill the gap. The probability to have {\em no} such secondaries is called 
the gap survival factor $S^2$.

We consider the physical origin of the $S^2$ suppression and describe the main 
elements of  $S^2$  calculation. We emphasize that actually this is not the 
common (constant) suppression factor. The value of $S^2$ depends on the 
particular subprocess and on the final state kinematics. In impact 
parameter space, $b$, the suppression $S^2(b)$ is very strong at small 
$b$, while at large $b$ we have $S^2\to 1$.
This leads to the dependence of $S^2$ on  transverse momenta of 
the outgoing protons.

An interesting situation is observed for the photon-mediated reactions.
The point is that the photon-parton distribution contains three different 
elements. One, which originates from `elastic' radiation ($p\to p+\gamma$),
corresponds to rather large $b$ where $S^2$ is close to 1. The second component arises from proton dissociation ($p\to N^*+\gamma$), which 
occupies smaller values of $b$. Finally, there are photons which are emitted from
quarks during the DGLAP evolution. This last part corresponds to small impact 
parameters and is suppressed most strongly by $S^2$. Therefore in the
calculation of CEP (or LRG) photon-induced cross sections we cannot 
use the usual inclusive photon PDF (which is known now to good 
accuracy~\cite{LUXqed}), but instead have to consider each component separately,
accounting for its own $S^2$ suppression.

\section*{Acknowledgements}

We thank Lucian Harland-Lang for valuable discussions and for research collaborations on many of the topics covered in this review. We thank Sergey Ostapchenko for useful discussions. VAK acknowledges  support from a Royal Society of Edinburgh  Auber award. MGR thanks the IPPP of Durham University for hospitality.

V.~A.~Khoze, A.~D.~Martin and M.~G.~Ryskin,
  Phys.\ Rev.\ D {\bf 96}, no. 3, 034018 (2017)
  [arXiv:1705.03685 [hep-ph]].

\bibitem{J-psi}
M.~G.~Ryskin,
  Z.\ Phys.\ C {\bf 57}, 89 (1993).
\bibitem{jones}
S.~P.~Jones, A.~D.~Martin, M.~G.~Ryskin and T.~Teubner,
  J.\ Phys.\ G {\bf 44}, no. 3, 03LT01 (2017)
  [arXiv:1611.03711 [hep-ph]].

\bibitem{Albrow:2010yb} M.~G. Albrow, T.~D. Coughlin and J.~R. Forshaw, 
{\em Prog.Part.Nucl.Phys.} {\bf  65}, 149  (2010), [arXiv:1006.1289 [hep-ph]]

\bibitem{Harland-Lang:2014lxa} 
  L.~A.~Harland-Lang, V.~A.~Khoze, M.~G.~Ryskin and W.~J.~Stirling,
  Int.\ J.\ Mod.\ Phys.\ A {\bf 29}, 1430031 (2014)
  [arXiv:1405.0018 [hep-ph]].

\bibitem{Harland-Lang:2015eqa} 
  L.~A.~Harland-Lang, V.~A.~Khoze and M.~G.~Ryskin,
  Int.\ J.\ Mod.\ Phys.\ A {\bf 30}, 1542013 (2015).

\bibitem{HarlandLang:2010ep}
L.~A. Harland-Lang, V.~A. Khoze, M.~G. Ryskin and W.~J. Stirling, {\em
  Eur.Phys.J.} {\bf C69}, 179  (2010),
 [arXiv:1005.0695 [hep-ph].

\bibitem{K_WMR} 
M.~A.~Kimber, A.~D.~Martin and M.~G.~Ryskin,
  Eur.\ Phys.\ J.\ C {\bf 12}, 655 (2000)
  [hep-ph/9911379];\\
A.~D.~Martin, M.~G.~Ryskin and G.~Watt,
  Eur.\ Phys.\ J.\ C {\bf 66}, 163 (2010)
  [arXiv:0909.5529 [hep-ph]].

\bibitem{Belitsky:2005qn}
A.~V. Belitsky and A.~V. Radyushkin, {\em Phys.Rept.} {\bf 418}, 1  (2005),
  [arXiv:hep-ph/0504030[hep-ph]].

\bibitem{Shuvaev:1999ce}
A.~Shuvaev, K.~J. Golec-Biernat, A.~D. Martin and M.~G. Ryskin, {\em Phys.Rev.}
  {\bf D60},   014015  (1999),
 {{\ttfamily arXiv:hep-ph/9902410 [hep-ph]}}.

\bibitem{Harland-Lang:2013xba} 
  L.~A.~Harland-Lang,
  Phys.\ Rev.\ D {\bf 88}, no. 3, 034029 (2013)
  [arXiv:1306.6661 [hep-ph]].

\bibitem{Cudell:1995ki} %
  J.~R.~Cudell and O.~F.~Hernandez,
  Nucl.\ Phys.\ B {\bf 471}, 471 (1996).

\bibitem{Cox:2001uq} 
  B.~Cox, J.~R.~Forshaw and B.~Heinemann,
  Phys.\ Lett.\ B {\bf 540}, 263 (2002)
  [hep-ph/0110173].
\bibitem{PR}
 M.~Boonekamp, A.~De Roeck, R.~B.~Peschanski and C.~Royon,
  Phys.\ Lett.\ B {\bf 550}, 93 (2002),
  [hep-ph/0205332].
\bibitem{Boonekamp:2004nu} 
  M.~Boonekamp, R.~B.~Peschanski and C.~Royon,
  Phys.\ Lett.\ B {\bf 598}, 243 (2004)
  [hep-ph/0406061].




\bibitem{Gluck94}
M.~Gluck, E.~Reya and A.~Vogt, {\em Z.Phys.} {\bf C67}, 433  (1995).

\bibitem{Martin:2009iq}
A.~D. Martin, W.~J. Stirling, R.~S. Thorne and G.~Watt, {\em Eur.Phys.J.} {\bf
  C63}, 189  (2009), [arXiv:0901.0002 [hep-ph]].

\bibitem{Pumplin:2002vw}
J.~Pumplin, D.~R. Stump, J.~Huston, H.~L. Lai, P.~M. Nadolsky {\em et~al.},
  {\em JHEP} {\bf 0207},   012  (2002)
[hep-ph/0201195].

\bibitem{Martin:1999ww}
A.~D. Martin, R.~Roberts, W.~J. Stirling and R.~S. Thorne, {\em Eur.Phys.J.}
  {\bf C14}, 133  (2000), [arXiv:hep-ph/9907231 [hep-ph]].
\bibitem{Khoze:2000mw} 
  V.~A.~Khoze, A.~D.~Martin and M.~G.~Ryskin,
  hep-ph/0006005.

\bibitem{Khoze:2000jm} 
  V.~A.~Khoze, A.~D.~Martin and M.~G.~Ryskin,
  Eur.\ Phys.\ J.\ C {\bf 19}, 477 (2001)
  Erratum: [Eur.\ Phys.\ J.\ C {\bf 20}, 599 (2001)]
  [hep-ph/0011393].

\bibitem{Khoze:2006uj}
V.~A. Khoze, A.~D. Martin and M.~G. Ryskin, {\em JHEP} {\bf 0605},   036
  (2006),  [arXiv:hep-ph/0602247 [hep-ph]].
\bibitem{Gotsman:2012rq}
E.~Gotsman, E.~Levin and U.~Maor, {\em Phys.Rev.} {\bf D85},   094007  (2012),
 [arXiv:1203.2419 [hep-ph]].


\bibitem{Gotsman:2014pwa}
E.~Gotsman, E.~Levin and U.~Maor, Int.\ J.\ Mod.\ Phys.\ A {\bf 30}, no. 08, 1542005 (2015)  [arXiv:1403.4531 [hep-ph]].

\bibitem{Khoze:2013jsa}
V.~A. Khoze, A.~D. Martin and M.~G. Ryskin, {\em Eur.Phys.J.} {\bf C74},   2756
   (2014), [arXiv:1312.3851 [hep-ph]].

\bibitem{Khoze:2002nf}
V.~A. Khoze, A.~D. Martin and M.~G. Ryskin, {\em Eur.Phys.J.} {\bf C24}, 581
  (2002), [arXiv:hep-ph/0203122 [hep-ph]].

\bibitem{Ryskin:2011qe}
M.~G. Ryskin, A.~D. Martin and V.~A. Khoze, {\em Eur.Phys.J.} {\bf C71},   1617
   (2011),[arXiv:1102.2844 [hep-ph]].


\bibitem{Watt}
  G.~Watt and H.~Kowalski,
  Phys.\ Rev.\ D {\bf 78}, 014016 (2008)
  [arXiv:0712.2670 [hep-ph]].

\bibitem{Ryskin:2009tj}
M.~G. Ryskin, A.~D. Martin and V.~A. Khoze, {\em Eur.Phys.J.} {\bf C60}, 249
  (2009),  [arXiv:0812.2407 [hep-ph]].

\bibitem{Ostap2} Sergey Ostapchenko, Marcus Bleicher, arXiv:1712.09695.
 
\bibitem{Kaidalov:1973tc} 
  A.~B.~Kaidalov, V.~A.~Khoze, Y.~F.~Pirogov and N.~L.~Ter-Isaakyan,
  Phys.\ Lett.\  {\bf 45B}, 493 (1973).

\bibitem{Field:1974fg} 
  R.~D.~Field and G.~C.~Fox,
  Nucl.\ Phys.\ B {\bf 80}, 367 (1974).

\bibitem{Luna:2008pp}
  E.~G.~S.~Luna, V.~A.~Khoze, A.~D.~Martin and M.~G.~Ryskin,
  Eur.\ Phys.\ J.\ C {\bf 59} (2009) 1
  [arXiv:0807.4115 [hep-ph]].

\bibitem{Harland-Lang:2013dia} 
  L.~A.~Harland-Lang, V.~A.~Khoze and M.~G.~Ryskin,
  Eur.\ Phys.\ J.\ C {\bf 74}, 2848 (2014)
  [arXiv:1312.4553 [hep-ph]].

\bibitem{Petrov:2009wr}
  V.~A.~Petrov, R.~A.~Ryutin and A.~E.~Sobol,
  Eur.\ Phys.\ J.\ C {\bf 65} (2010) 637
  [arXiv:0906.5309 [hep-ph]].

\bibitem{Ryutin:2016hyi} 
  R.~A.~Ryutin,
  Eur.\ Phys.\ J.\ C {\bf 77}, no. 2, 114 (2017)
  [arXiv:1612.03418 [hep-ph]].


\bibitem{Bjorken:1992er}
J.~Bjorken, {\em Phys.Rev.} {\bf D47}, 101  (1993).
\bibitem{Khoze:2003af}
  V.~A.~Khoze, W.~J.~Stirling and P.~H.~Williams,
  Eur.\ Phys.\ J.\ C {\bf 31} (2003) 91,[hep-ph/0307292].

\bibitem{Zep}
 H.~Chehime and D.~Zeppenfeld,
  Phys.\ Rev.\ D {\bf 47}, 3898 (1993);\\
D.~L.~Rainwater and D.~Zeppenfeld,
  JHEP {\bf 9712}, 005 (1997),
  [hep-ph/9712271].


\bibitem{Harland-Lang:2015cta} 
  L.~A.~Harland-Lang, V.~A.~Khoze and M.~G.~Ryskin,
  Eur.\ Phys.\ J.\ C {\bf 76}, no. 1, 9 (2016)
  [arXiv:1508.02718 [hep-ph]].

\bibitem{SuperCHIC}
The SuperCHIC code and documentation are available at {\tt
  http://projects.hepforge.org/superchic/}.

\bibitem{Harland-Lang:2017cax} 
  L.~A.~Harland-Lang, V.~A.~Khoze and M.~G.~Ryskin,
  arXiv:1709.00176 [hep-ph].


\bibitem{Piotrzkowski:2000rx} 
  K.~Piotrzkowski,
  Phys.\ Rev.\ D {\bf 63}, 071502 (2001)
  [hep-ex/0009065].

\bibitem{Pierzchala:2008xc}
  T.~Pierzchala and K.~Piotrzkowski,
Nucl.\ Phys.\ Proc.\ Suppl.\  {\bf 179-180} (2008) 257,
  [arXiv:0807.1121 [hep-ph]].

\bibitem{Chapon:2009hh} 
  E.~Chapon, C.~Royon and O.~Kepka,
  Phys.\ Rev.\ D {\bf 81}, 074003 (2010)
  [arXiv:0912.5161 [hep-ph]].

\bibitem{deFavereaudeJeneret:2009db} 
  J.~de Favereau de Jeneret {\it et al.},
  arXiv:0908.2020 [hep-ph].

\bibitem{Fichet:2013gsa} 
  S.~Fichet, G.~von Gersdorff, O.~Kepka, B.~Lenzi, C.~Royon and M.~Saimpert,
  Phys.\ Rev.\ D {\bf 89}, 114004 (2014).
\bibitem{Fichet:2014uka} 
  S.~Fichet, G.~von Gersdorff, B.~Lenzi, C.~Royon and M.~Saimpert,
  JHEP {\bf 1502}, 165 (2015),
  [arXiv:1411.6629 [hep-ph]].

\bibitem{Khoze:2017igg} 
  V.~A.~Khoze, A.~D.~Martin and M.~G.~Ryskin,
  J.\ Phys.\ G {\bf 44}, no. 5, 055002 (2017).

\bibitem{LUXqed}
  A.~Manohar, P.~Nason, G.~P.~Salam and G.~Zanderighi,
  Phys.\ Rev.\ Lett.\  {\bf 117}, no. 24, 242002 (2016)
  [arXiv:1607.04266 [hep-ph]];
  arXiv:1708.01256 [hep-ph].

\bibitem{MGRADM} A.~D.~Martin and M.~G.~Ryskin, Eur.Phys.J. C {\bf 74}, 3040 (2014) [arXiv:1406.2118 [hep-ph]]

\bibitem{Harland-Lang:2016apc} 
  L.~A.~Harland-Lang, V.~A.~Khoze and M.~G.~Ryskin,
  Eur.\ Phys.\ J.\ C {\bf 76}, no. 5, 255 (2016)
  [arXiv:1601.03772 [hep-ph]].

\bibitem{Harland-Lang:2016kog} 
  L.~A.~Harland-Lang, V.~A.~Khoze and M.~G.~Ryskin,
 Phys.\ Rev.\ D {\bf 94}, no. 7, 074008 (2016)
  [arXiv:1607.04635 [hep-ph]].

\bibitem{Chatrchyan:2011ci} 
  S.~Chatrchyan {\it et al.} [CMS Collaboration],
  JHEP {\bf 1201}, 052 (2012)
  [arXiv:1111.5536 [hep-ex]].

\bibitem{Chatrchyan:2012tv}
  S.~Chatrchyan {\it et al.} [CMS Collaboration],
  JHEP {\bf 1211}, 080 (2012)
  [arXiv:1209.1666 [hep-ex]].

\bibitem{Abbas:2013oua} 
  E.~Abbas {\it et al.} [ALICE Collaboration],
  Eur.\ Phys.\ J.\ C {\bf 73}, no. 11, 2617 (2013)
  [arXiv:1305.1467 [nucl-ex]].

\bibitem{Aad:2015bwa} 
  G.~Aad {\it et al.} [ATLAS Collaboration],
  Phys.\ Lett.\ B {\bf 749}, 242 (2015)
  [arXiv:1506.07098 [hep-ex]].

\bibitem{Aaboud:2017oiq}
  M.~Aaboud {\it et al.} [ATLAS Collaboration],
  arXiv:1708.04053 [hep-ex].

\bibitem{Chatrchyan:2013akv} 
  S.~Chatrchyan {\it et al.} [CMS Collaboration],
  JHEP {\bf 1307}, 116 (2013)
  [arXiv:1305.5596 [hep-ex]].

\bibitem{Aaboud:2016dkv}
  M.~Aaboud {\it et al.} [ATLAS Collaboration],
  Phys.\ Rev.\ D {\bf 94} (2016) no.3,  032011
  [arXiv:1607.03745 [hep-ex]].

\bibitem{Khachatryan:2016mud}
  V.~Khachatryan {\it et al.} [CMS Collaboration],
  JHEP {\bf 1608} (2016) 119
  [arXiv:1604.04464 [hep-ex]].

\bibitem{Tasevsky:2017wne} 
  M.~Tasevsky [ATLAS Collaboration],
  arXiv:1703.10472 [hep-ex].

\bibitem{Aaboud:2017oiq}
  M.~Aaboud {\it et al.} [ATLAS Collaboration],
  arXiv:1708.04053 [hep-ex].

\bibitem{Chapon:2009hh} 
  E.~Chapon, C.~Royon and O.~Kepka,
  Phys.\ Rev.\ D {\bf 81}, 074003 (2010)
  [arXiv:0912.5161 [hep-ph]].

\bibitem{CMS:2017uey}
  CMS Collaboration {\it et al.} [CMS and TOTEM Collaborations],
  CMS-PAS-PPS-17-001, TOTEM-NOTE-2017-003.

\bibitem{Hollar:2017njv} 
  J.~Hollar {\it et al.} [TOTEM Collaboration],
  arXiv:1709.02985 [hep-ex].

\bibitem{Aaboud:2017bwk} 
  M.~Aaboud {\it et al.} [ATLAS Collaboration],
  Nature Phys.\  {\bf 13}, no. 9, 852 (2017)

\bibitem{Khoze:2000db}
 V.~A.~Khoze, A.~D.~Martin, R.~Orava and M.~G.~Ryskin,
  Eur.\ Phys.\ J.\ C {\bf 19}, 313 (2001)
  [hep-ph/0010163].

\bibitem{Krasny:2006xg} 
  M.~W.~Krasny, J.~Chwastowski and K.~Slowikowski,
  Nucl.\ Instrum.\ Meth.\ A {\bf 584}, 42 (2008)
  [hep-ex/0610052].
  
 \bibitem{LHCbJ} R. Aaij  {\it et al.} [LHCb Collaboration], CERN-LHCb-Conf-2016-007 (2016).

 \bibitem{jones2} S.~P.~Jones, A.~D.~Martin, M.~G.~Ryskin and T.~Teubner, JHEP {\bf 1311}, 085 (2013) [arXiv:1307.7099 [hep-ph]]; 
  Eur. Phys. J. {\bf C76}, 633 (2016)   [arXiv:1610.02272 [hep-ph]].

\bibitem{Budnev:1974de} 
  V.~M.~Budnev, I.~F.~Ginzburg, G.~V.~Meledin and V.~G.~Serbo,
  Phys.\ Rept.\  {\bf 15}, 181 (1975).
  [arXiv:1702.01625 [hep-ex]].



\bibitem{Khoze:2002dc} 
  V.~A.~Khoze, A.~D.~Martin and M.~G.~Ryskin,
  Eur.\ Phys.\ J.\ C {\bf 24}, 459 (2002)
  [hep-ph/0201301].

\bibitem{HarlandLang:2012qz} 
  L.~A.~Harland-Lang, V.~A.~Khoze, M.~G.~Ryskin and W.~J.~Stirling,
  Eur.\ Phys.\ J.\ C {\bf 72}, 2110 (2012)
  [arXiv:1204.4803 [hep-ph]].
  
\bibitem{Vermaseren:1982cz} 
  J.~A.~M.~Vermaseren,
  Nucl.\ Phys.\ B {\bf 229}, 347 (1983).

\bibitem{Baranov:1991yq} 
  S.~P.~Baranov, O.~Duenger, H.~Shooshtari and J.~A.~M.~Vermaseren,
\newblock {Hamburg 1991, Proceedings, Physics at HERA, vol. 3, 1478-1482. (see
  HIGH ENERGY PHYSICS INDEX 30 (1992) No. 12988)}  (1991)).

\bibitem{Dyndal:2014yea}
  M.~Dyndal and L.~Schoeffel,
  Phys.\ Lett.\ B {\bf 741} (2015) 66
  [arXiv:1410.2983 [hep-ph]].
  
\bibitem{dEnterria:2013zqi}
  D.~d'Enterria and G.~G.~da Silveira,
  Phys.\ Rev.\ Lett.\  {\bf 111} (2013) 080405;
   Erratum: [Phys.\ Rev.\ Lett.\  {\bf 116} (2016) no.12,  129901]
  [arXiv:1305.7142 [hep-ph]].
   \bibitem{Klusek-Gawenda:2016euz} 
  M.~Klusek-Gawenda, P.~Lebiedowicz and A.~Szczurek,
  Phys.\ Rev.\ C {\bf 93}, no. 4, 044907 (2016)
  [arXiv:1601.07001 [nucl-th]].

\bibitem{Ellis:2017edi}
  J.~Ellis, N.~E.~Mavromatos and T.~You,
  Phys.\ Rev.\ Lett.\  {\bf 118} (2017) no.26,  261802
  [arXiv:1703.08450 [hep-ph]].

\bibitem{Knapen:2017ebd}
  S.~Knapen, T.~Lin, H.~K.~Lou and T.~Melia,
  arXiv:1709.07110 [hep-ph].


\bibitem{Albrow:2014lta} 
  M.~Albrow {\it et al.} [CMS and LHCb Collaborations and FSC Team and HERSCHEL Team],
  Int.\ J.\ Mod.\ Phys.\ A {\bf 29}, no. 28, 1446018 (2014).

\bibitem{McNulty}
  K.~Carvalho Akiba {\it et al.},
  arXiv:1801.04281 [physics.ins-det].

\bibitem{Schicker:2014wvk}
  R.~Schicker,
  Int.\ J.\ Mod.\ Phys.\ A {\bf 29} (2014) 1446015
  [arXiv:1411.1283 [hep-ex]];\\
A.~Villatoro Tello [ALICE Collaboration],
  AIP Conf.\ Proc.\  {\bf 1819} (2017) no.1,  040020.

\bibitem{Albrow:2008az}
  M.~Albrow {\it et al.},
  JINST {\bf 4} (2009) P10001
  [arXiv:0811.0120 [hep-ex]].

\bibitem{Khoze:2149} 
  V.~A.~Khoze, A.~D.~Martin, M.~G.~Ryskin,
    Eur.\ Phys.\ J.\ C {\bf 73}, 2503 (2013) [arXiv:1306.2149 [hep-ph]].
\bibitem{Khoze:2010by} 
  V.~A.~Khoze, F.~Krauss, A.~D.~Martin, M.~G.~Ryskin and K.~C.~Zapp,
  Eur.\ Phys.\ J.\ C {\bf 69}, 85 (2010)
  [arXiv:1005.4839 [hep-ph]].

\bibitem{Edin:1995gi}
   A.~Edin, G.~Ingelman and J.~Rathsman,
   Phys.\ Lett.\ B {\bf 366}, 371 (1996)
   [hep-ph/9508386].


\bibitem{Jung:1993gf}
   H.~Jung,
   Comput.\ Phys.\ Commun.\  {\bf 86}, 147 (1995).

\bibitem{Cox:2000jt}
   B.~E.~Cox and J.~R.~Forshaw,
   Comput.\ Phys.\ Commun.\  {\bf 144} (2002) 104
   [hep-ph/0010303].

\bibitem{Boonekamp:2011ky}
   M.~Boonekamp, A.~Dechambre, V.~Juranek, O.~Kepka, M.~Rangel, C.~Royon
and R.~Staszewski,
   arXiv:1102.2531 [hep-ph].
   
\bibitem{Monk:2005ji}
   J.~Monk and A.~Pilkington,
   Comput.\ Phys.\ Commun.\  {\bf 175}, 232 (2006)
   [hep-ph/0502077]. 

\bibitem{Early}
V.A. Khoze, A.D. Martin and M.G. Ryskin, Eur. Phys. J. {\bf C55}, 363 (2008)
[arXiv:0802.0177 [hep-ph]].


\end{document}